\documentclass[12pt,A4]{article}
\usepackage{amssymb, amsfonts, amsmath, amsthm, eurosym, mathabx, graphicx, subfig, srcltx, calc, epsfig, color, multirow}
\usepackage[multiple]{footmisc}
\usepackage{tablefootnote}
\usepackage{array}
\usepackage[round]{natbib}
\usepackage{thmtools}
\usepackage{thm-restate}
\usepackage{enumitem}
\usepackage{algorithm}
\usepackage{algpseudocode}
\usepackage{titlesec}

\usepackage{mathrsfs}
\usepackage{hyperref}

    \newtheorem{theorem}{\sc Theorem}

    \newtheorem{corollary}{\sc Corollary}

    \newtheorem{proposition}{\sc Proposition}

    \newcommand {\argmax}{\mbox{argmax}}

    \newcommand {\mbx}{\mathbf{x}}
    \newcommand {\mbp}{\mathbf{p}}

\usepackage{comment}

\usepackage{xcolor} 
\usepackage{tikz}
\usetikzlibrary{trees}
\usetikzlibrary{patterns}
\usepackage[english]{babel}

\titleformat{\section}
  {\normalfont\fontsize{15}{15}\bfseries}{\thesection}{1em}{}

\titleformat{\subsection}
  {\normalfont\fontsize{15}{15}\bfseries}{\thesubsection}{1em}{}
  
\begin{document}

\title{\textbf{Price Heterogeneity\\ as a Source of \\ Heterogeneous Demand%
\thanks{%
\linespread{1.1} \selectfont We thank Kwok Ping Tsang for helpful discussions. We also thank the seminar participants at U of Connecticut, CUHK-HKU-HKUST,  KAIST, UC Santa Cruz, U of Sussex, Purdue U, U of Pennsylvania, and U of Manitoba, along with the audiences of D-TEA 2025 (PSE) and RUD 2025 (Manchester).}}}
\author{John K.-H.  Quah\thanks{%
\linespread{1.1} \selectfont Department of Economics, National University of Singapore. E-mail: ecsqkhj@nus.edu.sg.} \and Gerelt Tserenjigmid\thanks{Department of Economics, University of California, Santa Cruz. E-mail: gtserenj@ucsc.edu.}}

\maketitle

\begin{abstract} 
\noindent We explore heterogeneous prices as a source of heterogeneous or stochastic demand. Heterogeneous prices could arise either because there is actual price variation among consumers or because consumers (mis)perceive prices differently.  Our main result says the following: if heterogeneous prices have a distribution among consumers that is (in a sense) stable across observations, then a model where consumers have a common utility function but face heterogeneous prices has precisely the same implications as a heterogeneous preference/random utility model (with no price heterogeneity).\vspace{0.05in}

\end{abstract}\vspace{-0.1in}

\section{Introduction}\label{intro}

\noindent Why is demand heterogeneous?  Two consumers with the same budget and facing the same prices would often choose to spend their money differently. The same consumer, when presented with the same budgetary constraints, may also choose different purchases at different times.  The most common and obvious way of modeling heterogeneous behavior across consumers (or stochastic behavior in the case of a single consumer) is to assume that demand is governed by a distribution of preferences (rather than a single preference).  When the distribution of preferences is \emph{stable} across multiple observations, this leads to a random utility model of stochastic choice (see, for example, \cite{mcfadden1990stochastic}, \cite{kitamura2018nonparametric}, and \cite{deb2023revealed}).  But preference heterogeneity is not the only source of variation in choice behavior.  More recently, the literature on stochastic choice has highlighted the presence of other sources of heterogeneous choice.  For example, choice could be stochastic due to random consideration/attention sets  (\cite{manzini2014stochastic}, \cite{brady2016menu}, \cite{cattaneo2020random}) or due to deliberate randomization (\cite{fudenberg2015stochastic}, \cite{cerreia2019deliberately}).  In this paper, we wish to highlight another reason why choices can be heterogeneous: the presence of heterogeneous prices across consumers.  Price heterogeneity could occur for at least two reasons.

Firstly, even when the economist uses only a single set of prices (for different goods) for all the consumers being studied, consumers {\em in fact} face different prices due to {\bf price dispersion} (\cite{stigler1961economics}), which arises because they live in different locations and/or shop at different stores, and it is not always worthwhile for them to search for the cheapest price.\footnote{\linespread{1.1} \selectfont For empirical evidence on price dispersion, see \cite{sorensen2000equilibrium}, \cite{brown2002does}, \cite{baye2004price}, \cite{van2010has}, and \cite{kaplan2019relative}.}  Related to this, empirical demand analysis often involves the aggregation of highly granular demand information into broad categories (such as `food' or `clothing'); the price index constructed for a given category of goods is not tailored to each consumer and so may not accurately represent the prices encountered by a given consumer for goods in that category.

Secondly, even when consumers face the same prices, they may still (mis)perceive prices differently, leading to different demand behavior. {\bf Price misperception} could occur because of {\em inattention to prices} (\cite{chetty2009salience}, \cite{gabaix2014sparsity},  \cite{matvejka2015rigid}) or because consumers over- or under-estimate their usage of a prepaid good or service (\cite{dellavigna2006paying}).  A related phenomenon is that the demand for a product could be affected by its {\em reference price} (see \cite{koszegi2006model}, \cite{heidhues2008competition}, \cite{gennaioli2013salience}), with the same price being perceived as high or low depending on the reference price.  The variation in reference prices across consumers then leads to heterogeneous demand behavior.\footnote{\linespread{1.1} \selectfont  There is a large literature in marketing on reference-price purchasing. See \cite{mazumdar2005reference} for a survey.}


In our formal analysis, we assume that the analyst has information on the expenditure allocations (across $K$ goods) of $N$ consumers at different price indices. Formally, a data set with $T$ observations can be represented by
$$\mathcal{E}=\big\{(\mathbf{e}^{i, t})_{i \in N}, \overline{\mathbf{p}}^t\big\}_{t\in T},$$
where $\mathbf{e}^{i, t}=(e^{i,t}_k)_{k\in K}$ is the expenditure allocation of consumer $i$ and $\overline{\mathbf{p}}^{t}=(\overline{p}^t_k)_{k\in K}$ are the prices of the $K$ goods at observation $t$.\footnote{ Note that $T$ (similarly $N$ and $K$) denotes both a set and the number of elements in the set.}  If there is no price heterogeneity, then consumer $i$'s consumption at observation $t$ is the bundle $\mathbf{x}^{i,t}$, where the $k$th entry, $x_k^{i,t}=e^{i,t}_k/\overline{p}^{t}_k$.  We investigate the implications of price heterogeneity in two different models of demand.\vspace{0.1in}


\noindent {\bf Constrained utility-maximization.}\, Afriat's Theorem (\cite{afriat1967construction}, \cite{varian1982nonparametric}) allows us to check whether a consumer's demand behavior is consistent with the maximization of a utility function subject to a budget constraint, i.e., whether, for a consumer $i$, there is a utility function $u^i$ such that $u^i(\mathbf{x}^{i,t})\geq u^i(\mathbf{x})$ for all $\mathbf{x}$ satisfying $\overline{\mathbf{p}}^t\cdot \mathbf{x}\leq \sum_{k\in K}e^{i,t}_k$. The theorem states that this holds if and only if the implied consumption bundles of consumer $i$, $\{\mathbf{x}^{i,t}\}_{t\in T}$, along with the observed prices $\{\overline{\mathbf{p}}^t\}_{t\in T}$, satisfy a property called the generalized axiom of revealed preference (or GARP, for short).  If this holds for each agent $i$, then we say that $\cal E$ can be {\em rationalized by heterogeneous preferences}.

The problem we pose is different.  Instead of allowing consumers to have different utility functions, we require them to have the {\em same} utility function but allow each consumer to face idiosyncratic (whether real or perceived) prices $\mathbf{p}^{i,t}$.  We say that the data set $\cal E$ can be {\em rationalized by heterogeneous prices} if there is a common utility function $u$ such that consumer $i$'s expenditure allocations at different observations maximize $u$, given the idiosyncratic prices $\{\mathbf{p}^{i,t}\}_{t\in T}$ encountered by consumer $i$; formally, we require  $u(\widehat{\mathbf{x}}^{i,t})\geq u(\mathbf{x})$, where $\widehat{x}^{i,t}_k=e^{i,t}_k/p^{i,t}_k$ and $\mathbf{x}$ satisfies $\mathbf{p}^{i,t}\cdot\mathbf{x}\leq \sum_{k\in K}e^{i,t}$. 

The interpretation of the bundles $\{\widehat{\mathbf{x}}^{i,t}\}_{t\in T}$ depends on the interpretation we give to $\{\mathbf{p}^{i,t}\}$.
\begin{itemize}
\item  If we interpret $\{\mathbf{p}^{i,t}\}_{t\in T}$ as the true prices faced by consumer $i$, then $\widehat{\mathbf{x}}^{i,t}$ is the true bundle purchased by this consumer at observation $t$; this bundle is not observed by the researcher, who only observes $i$'s expenditure allocation $\mathbf{e}^{i, t}$ and the price index $\overline{\mathbf{p}}^t$.  Under this interpretation, if $\cal E$ can be rationalized by heterogeneous prices, then all consumers share the same utility function and each of them is rational, in the sense that there are idiosyncratic prices at which each consumer's observed expenditure allocations are optimal. 
\item  If we interpret $\{\mathbf{p}^{i,t}\}_{t\in T}$ as consumer $i$'s misperceived prices, then $\widehat{\mathbf{x}}^{i,t}$ is the bundle that this consumer {\em thinks} she is consuming when making her expenditure allocation decision across goods. We are assuming that the expenditure allocation $\mathbf{e}^{i,t}$ is the one consumer $i$ implements so that the consumer's actual consumption at observation $t$ is $\mathbf{x}^{i,t}$.  Under this interpretation, if $\cal E$ can be rationalized by heterogeneous prices, then all consumers share the same utility function and each consumer's expenditure allocations are also rational, once her misperception of prices is taken into account.
\end{itemize}
 
Our notion of rationalization, in itself, is too permissive to lead to meaningful implications on data; what is needed are sensible restrictions on how heterogeneous prices are distributed across consumers.  In Section \ref{result1}, we require the heterogeneous prices to be {\em correct on average}, in the sense that, at each observation $t$ and for each good $k$, we have $(\sum_{i\in N}p^{i,t}_k)/N=\overline{p}_k^t$. In this case, \textbf{we show that {\em any} data set $\cal E$ can be rationalized by heterogeneous prices}. In particular, price heterogeneity, in this sense, is more powerful than preference heterogeneity in its ability to explain observed behavior, since each consumer's expenditure allocation (along with the prevailing price indices) must satisfy GARP in the latter model.

In Section \ref{stable}, we impose a stronger restriction on price heterogeneity.  We assume that price heterogeneity has a {\em stable structure} across observations. This assumption requires that the distribution of prices for good $k$ (across consumers) at some observation $t$ is the same as the distribution of prices for good $k$ at observation $t'$, after normalizing by the average price of $k$ at each observation.  When this condition is imposed on the price distributions, we find that stable price heterogeneity leads to the {\em same} restrictions on $\cal E$ as preference heterogeneity.  In other words, \textbf{\em a data set $\cal E$ can be rationalized by heterogeneous preferences if and only if it can be rationalized by heterogeneous prices with a stable structure}.

The above equivalence result leads to two additional equivalence results that support two common practices in empirical work. First, we can show that \textbf{\emph{a data set $\cal E$ can be rationalized by heterogeneous preferences if and only if it can be rationalized by heterogeneous preferences and any heterogeneous prices with a stable structure}}. Since both preference and price heterogeneity are prevalent, the latter model seems more realistic, yet it is common practice in empirical work to assume price homogeneity. Our result supports this practice of ignoring price heterogeneity. 

Second, we show that \textbf{if $\cal E$ can be rationalized by heterogeneous preferences, then all the preferences can be drawn from a family of utility functions related to each other via equivalence scales.} Equivalence scales are widely used in empirical models of demand to capture (among other things) the effects of household composition (see, for example, \cite{lewbel1989household}, \cite{blundell1991information}, \cite{deaton1997analysis}, and \cite{lewbel2017unobserved}) and in models of market demand they can be used to endow a parametric structure to preference heterogeneity (see, for example,  \cite{grandmont1992transformations} and \cite{quah1997law}). Hence, our result supports the practice of using this restrictive class of utility functions.


\medskip
\noindent {\bf Augmented utility-maximization.}\, In Section \ref{sec-augment}, we study the impact of price heterogeneity in a second model of demand. Instead of a consumer who maximizes utility subject to a budget constraint, we posit a consumer who maximizes a utility function where expenditure enters directly. Formally, the utility of a bundle $\mathbf{x}$ acquired at cost $e$ is $V(\mathbf{x},-e)$, where $V$ is strictly increasing in the last argument (in other words, higher expenditure gives strictly lower utility). Following \cite{deb2023revealed}, we refer to $V$ as an {\em (expenditure-)augmented utility function}.  As a special case, $V$ can have the familiar quasilinear form, where $V(x,-e)=U(x)-e$ for some function $U$. 

Setting aside price misperception for now, at prices $\mathbf{p}$ the consumer chooses the bundle $\mathbf{x}$ to maximize $V(\mathbf{x},-\mathbf{p}\cdot \mathbf{x})$. Notice that she is deterred from buying an arbitrarily large bundle because of the disutility that higher expenditure incurs.  An augmented utility model is particularly suited for the study of demand under price misperception because it allows for any bundle to be purchased, even a bundle chosen under misperceived prices; on the other hand, in a model with constrained optimization, such a bundle will not be implementable if it violates the budget constraint.\footnote{\linespread{1.1} \selectfont Our solution to this problem is to assume that the consumer always implements her {\em expenditure} allocation decision. For another solution to this problem, see \cite{gabaix2014sparsity}.}  Given a data set $\cal E$, we could ask whether $\cal E$ is rationalizable by heterogeneous augmented utility functions, assuming that (at each observation) prices are homogeneous across consumers. \cite{deb2023revealed} provide a characterization of those data sets $\cal E$ which can be rationalized in this sense.  

We are interested in investigating the conditions under which $\cal E$ can be rationalized by a model where all consumers have the same augmented utility function but associates to each bundle $\mathbf{x}$ an expenditure level that is not simply $\overline{\mathbf{p}}^t\cdot \mathbf{x}$. We allow for heterogeneous prices $\{\mathbf{p}^{i,t}\}_{t\in T}$, with consumer $i$ (mentally) associating with the bundle $\mathbf{x}$ a {\em behavioral expenditure level}\, $\phi (\overline{\mathbf{p}}^t\cdot \mathbf{x},\mathbf{p}^{i,t}\cdot\mathbf{x})$. Note that this value depends on the true cost of the bundle, $\overline{\mathbf{p}}^t\cdot \mathbf{x}$, as well as on $\mathbf{p}^{i,t}\cdot\mathbf{x}$. Different functional restrictions on $\phi$ allow the behavioral expenditure level to capture different notions of price (and, hence, expenditure) misperception, including price inattention and reference prices; in the former case, $\mathbf{p}^{i,t}$ are consumer $i$'s idiosyncratic perceived prices and in the latter, $\mathbf{p}^{i,t}$ are consumer $i$'s idiosyncratic reference prices.  The data set $\cal E$ is said to be rationalized by an augmented utility function $V$ and heterogeneous prices if $\mathbf{x}^{i,t}$ (consumer $i$'s actual consumption bundle given the true prices $\overline{\mathbf{p}}^t$) maximizes $V(\mathbf{x},-\phi(\overline{\mathbf{p}}^t\cdot\mathbf{x},\mathbf{p}^{i,t}\cdot \mathbf{x}))$ at each observation $t$.  We show that, {\em provided $\phi$ satisfies some mild conditions, \textbf{$\cal E$ can be rationalized by an augmented utility function and heterogeneous prices with a stable structure if and only if it can be rationalized by heterogeneous augmented utility functions}}. Loosely speaking, heterogeneity in price misperception has the same observable implications as heterogeneity in preferences when consumers maximize augmented utility.

\section{Model and basic definitions}\label{model}

There are $N$ consumers who purchase bundles of $K$ goods.  We denote by $e^{i, t}_k$ the amount that consumer $i$ spends on good $k$ at observation $t$.  We assume that $e^{i,t}_k\geq 0$ and $\sum_{k\in K}e^{i,t}_k>0$, i.e., consumer $i$'s total expenditure is always strictly positive, though expenditure on a given good may equal zero.   These expenditure allocations are observed by an analyst. Consumer $i$ at observation $t$ bases his expenditure allocation decision on the prices $\mathbf{p}^{i,t}=(p^{i, t}_k)_{k\in K}$.  Notice that these prices can be idiosyncratic; we assume that they are {\em not} observed by the analyst.\footnote{\linespread{1.1} \selectfont For example, in the panel study of income dynamics (PSID) survey, which is one of the most applied data sets in economics, households report their expenditures but not the individual prices that they face.} However, the analyst does observe $\overline{p}^t_k$, which is an aggregate (or index) of the prices of good $k$ at observation $t$.  Assuming that the analyst makes $T<\infty$ observations, we may denote the data set collected in the form
\begin{equation}\label{ourdata set}
\mathcal{E}=\big\{\big(\mathbf{e}^{i, t}\big)_{i\in N}, \overline{\mathbf{p}}^{t}\big\}_{t\in T},
\end{equation}
where $\mathbf{e}^{i,t}=(e^{i,t}_k)_{k\in K}$ and $\overline{\mathbf{p}}^t=(\overline{p}^t_k)_{k\in K}$.  (Notice that we have abused notation by using $T$ to denote both the number of observations and the set $\{1,2,\ldots,T\}$, and similarly with $K$ and $N$.) We assume that $p^{i,t}_k>0$ and $\overline{p}^t_k>0$, i.e., for every good and at every observation, the prices faced by the consumers and the price index are {\em strictly} positive.

There are two possible interpretations of the idiosyncratic prices:\\
\textbf{(1) Observational Errors}\;  The analyst's price index $\overline{\mathbf{p}}^t$ represents a simplification of a more complex environment; $\mathbf{p}^{i,t}$ are the true prices faced by consumer $i$ and they could be different from $\overline{\mathbf{p}}^t$ because of geographical variation in prices, variation in prices within a particular observation period $t$, or unobserved costs which are incurred by the consumer when purchasing a particular good. Price heterogeneity is also plausible when each product represents a broad category (such as `food' or `clothing'). In these situations, the data available to the observer takes the form of expenditure on a product category $k$ by each consumer and a price index for goods in that category. Simply deflating each consumer's expenditure by this index can be misleading if consumers are systematically different in the types of goods they purchase within category $k$.  \\
\textbf{(2) Price Misperception}\; Consumer $i$ makes his expenditure allocation decision based on an idiosyncratic perception of prices at observation $t$.  The prices $\mathbf{p}^{i,t}$ could be incorrect or misperceived, in the sense that they are different from the true prices $\overline{\mathbf{p}}^t$, which (unlike in the first interpretation) are perfectly observed by the analyst. Price misperception could occur because consumers are inattentive to prices (\cite{chetty2009salience}, \cite{gabaix2014sparsity},  \cite{matvejka2015rigid}), for example, failing to incorporate sales taxes or shipping charges.  Sometimes the phenomenon could also be exacerbated by firms engaging in price shrouding (\cite{gabaix2006shrouded}, \cite{brown2010shrouded}).  Phenomena, where people misperceive their usage of a prepaid good/service such as a gym membership, could also be considered as a form of price misperception (see \cite{dellavigna2006paying}).\footnote{\linespread{1.1} \selectfont Price misperception is a form of bounded rationality, but it is formally different from models (such as \cite{masatlioglu2012revealed} and \cite{manzini2014stochastic}) where agents only consider some of the options open to them. The `consideration set' in these models is always a subset of the budget set, whereas misperceived prices lead to a misperceived budget that is typically {\em not} a subset of the true budget set.}

We would like to find conditions under which the data set $\mathcal{E}$ has the following properties:\, (i) at each observation, the heterogeneous prices across consumers are consistent with the observed price indices in some sense and (ii) there is a single utility function that could explain the observed expenditure patterns. To explain this more precisely, some definitions are in order.

For given expenditure and price vectors $\mathbf{e}\in\mathbb{R}^K_+$ and $\mathbf{p}\in\mathbb{R}^K_{++}$, we denote the implied consumption bundle by $\mathbf{x}(\mathbf{e},\mathbf{p})$; formally, $\mathbf{x}(\mathbf{e}, \mathbf{p})\equiv \big(e_k/p_k\big)_{k\in K}.$ The set of bundles that cost as much as or less than $\mathbf{x}(\mathbf{e}, \mathbf{p})$ is given by the {\em budget set}
$$\mathscr{B}(\mathbf{p},m)= \{\mathbf{x}\in\mathbb{R}^{K}_{+}| \mathbf{p}\cdot \mathbf{x}\le m\},$$
where $m=\sum_{k\in K}e_k$ is the total expenditure.\bigskip

\noindent\textbf{Utility Maximization.}\, Let ${\cal P}=\{\mathbf{p}^{i,t}\}_{(i,t)\in N\times T}$ be a collection of heterogeneous prices.  We say that $\cal P$ \textbf{rationalizes} $\mathcal{E}=\big\{\big(\mathbf{e}^{i, t}\big)_{i\in N}, \overline{\mathbf{p}}^{t}\big\}_{t\in T}$ if there is a utility function $u:\mathbb{R}^K_{+}\to\mathbb{R}$ such that for each $(i, t)\in N\times T$,
\begin{equation} \label{budg-max}
u(\mathbf{x}(\mathbf{e}^{i,t}, \mathbf{p}^{i,t})) \ge u(\mathbf{x}) \text{ for any }\mathbf{x}\in \mathscr{B}\left(\mathbf{p}^{i,t}, m^{i,t}\right),
\end{equation}
\[ \mbox{where }\:\: m^{i, t}=\sum\nolimits_{k\in K} e^{i,t}_k.\]
In other words, there is a utility function (common to all consumers) such that each consumer's demand at observation $t$ (implied by the observed expenditure shares and the idiosyncratic prices) gives greater utility than any bundle that costs as much or less.  The data set $\cal E$ is \textbf{rationalizable with heterogeneous prices} if there are heterogeneous prices $\cal P$ that rationalize $\cal E$, typically subject to some consistency condition on $\cal P$ which we shall discuss later.

Notice that in this notion of rationalization, there is {\em heterogeneity in prices across consumers} but no heterogeneity in preferences.  The precise  interpretation of $\mathbf{x}(\mathbf{e}^{i,t}, \mathbf{p}^{i,t})$ depends on our interpretation of idiosyncratic prices.

In the first (Errors) interpretation of idiosyncratic prices, these prices are true prices and so $\mathbf{x}(\mathbf{e}^{i,t}, \mathbf{p}^{i,t})$ is the actual bundle purchased by the consumer, while (\ref{budg-max}) guarantees that each consumer is truly optimizing at the utility function $u$.

In the second (Misperceptions) interpretation, the bundle $\mathbf{x}(\mathbf{e}^{i,t}, \mathbf{p}^{i,t})$ is what consumer $i$ expects to be able to buy at his perceived prices and (\ref{budg-max}) guarantees that, with utility function $u$, consumer $i$ thinks that his expenditure allocation decision is optimal, given his perception of prices. Since our model assumes that the allocated expenditures are {\em observed}, we are assuming that the consumer implements the expenditure allocation decision, so that (with $\overline{\mathbf{p}}^t$ being the true prices) the actual level of good $k$ consumed by $i$ at observation $t$ is $e^{i,t}_k/\overline{p}_k^t$.  Notice that the expenditure allocation decision by the consumer will typically not be optimal given the true prices, and the consumer may even become aware of the true prices as he implements his expenditure allocation decision, but we are assuming that he does not re-optimize: he retains the expenditure allocation decision and simply adjusts by buying more or less of the good.\footnote{\linespread{1.1} \selectfont For another way of reconciling price misperception with a binding budget constraint see \cite{gabaix2014sparsity}.  In Section \ref{sec-augment}, we present a different model of consumer demand where there is no binding budget constraint and so even a choice under misperceived prices can be implemented exactly.}\medskip

\noindent\textbf{Price Consistency.}  To render this model interesting, there must be some way of disciplining the sort of price heterogeneity permitted by the model.  We do this via two notions of price consistency.

We call a function $W:\mathbb{R}^{N}_{++}\to \mathbb{R}_{++}$ a \textbf{price aggregator} if it is strictly increasing, continuous, and $W\big(p, \ldots, p\big)=p$ for any $p>0$. For example, the arithmetic average across all consumers
\begin{equation}\label{arith-ave}
W((p^i)_{i\in N})=\frac{\sum_{i\in N} p^i}{N}
\end{equation}
is a price aggregator.  We assume that, at each observation $t$ and for each good $k$, there is an aggregator function $W^t_k$ that is known to the analyst. In our first result (Proposition \ref{One}), we assume that the aggregator functions $\mathcal{W}=\{W^t_k\}_{k\in K, t\in T}$ satisfy one of the following regularity conditions:
\begin{enumerate}
\item[i)] $\lim_{p^i\to+\infty }W^t_k\big(p^i, (p^{j})_{j\neq i}\big)=+\infty$ for any $(p^{j})_{j\neq i}\in\mathbb{R}^{N-1}_{++}$ and $(t, k)\in T\times K$, or
\item[ii)] $\lim_{p^i\to 0}W^t_k\big(p^i, (p^{j})_{j\neq i}\big)=0$ for any $(p^{j})_{j\neq i}\in\mathbb{R}^{N-1}_{++}$ and $(t, k)\in T\times K$.
\end{enumerate}

Given the price aggregators $\mathcal{W}=\{W^t_k\}_{(k,t)\in K\times T}$, the idiosyncratic prices ${\cal P}=\{\mathbf{p}^{i, t}\}_{(i,t)\in N\times T}$ are \textbf{consistent in expectation} with the price indices $\{\overline{\mathbf{p}}^t\}_{t\in T}$ if, for any $(k, t)\in K\times T$,
\begin{equation}\label{idiosyn}
\overline{p}^t_k=W^t_k\big((p^{i, t}_k)_{i\in N}\big).
\end{equation}
In the Errors interpretation of idiosyncratic prices, $\cal W$ could be the actual formulae used in the calculation of the observed price indices, so that (\ref{idiosyn}) requires that the hypothesized $\cal P$ agrees on average with the aggregate price data.  In the Misperceptions interpretation, (\ref{idiosyn}) says that the consumers are on average correct in their price perceptions  (according to some suitable notion of average imposed by the modeler via $\cal W$).\footnote{\linespread{1.1} \selectfont Misperception restrictions of this type are not unique to our model; indeed, our notion is akin to the `rational expectations' assumption made, for example, in \cite{koszegi2006model, koszegi2007reference}.}

The second notion of price consistency we consider is to require $\cal P$ to have a \textbf{stable distribution} with respect to the price indices $\{\overline{\mathbf{p}}^t\}_{t\in T}$.  This means that there is $\lambda_{i,k}$ such that for any $t\in T$,
\begin{equation}
p^{i,t}_k=\lambda_{i,k}\,\overline{p}^t_k.
\end{equation}
For each good, the distribution of heterogeneous prices is stable across observations; indeed, if consumer $i$ assigns a price to $k$ that is 10\% above $\overline{p}^{t}_k$ at {\em some} observation $t$, then he will do the same at {\em every} observation.   It is clear that a stable price distribution will also be consistent in expectation under the following assumptions:\, (a) $W^t_k$ does not vary with $t$ and thus can be denoted by $W_k$; and (b) $W_k$ is homogeneous of degree 1, with the normalization that $W_k( (\lambda_{i,k})_{i\in N})=1$ for all $k$. Obviously, these conditions hold if the price aggregator is the arithmetic average (\ref{arith-ave}) for every good and every observation.\medskip

\noindent\textbf{Preference heterogeneity.}\, Of course, the standard way of imposing structure on a data set is to require that all agents face the same prices but allow them to have different preferences. The data set $\mathcal{E}=\{(\mathbf{e}^{i, t}, \overline{\mathbf{p}}^{t})\}_{(i, t)\in N\times T}$ is \textbf{rationalizable with heterogeneous preferences} if there are strictly increasing and continuous utility functions $u^i:\mathbb{R}^K_+\to\mathbb{R}$ such that for each $(i, t)\in N\times T$,
\begin{equation} \label{budg-max2}
u^i(\mathbf{x}(\mathbf{e}^{i,t}, \overline{\mathbf{p}}^{t})) \ge u^i(\mathbf{x}) \text{ for any }\mathbf{x}\in \mathscr{B}\left(\overline{\mathbf{p}}^{t}, m^{i,t}\right).
\end{equation}
\smallskip

\noindent\textbf{Cross-Sectional Data Environment.}\,  As we explore the implications of price heterogeneity in this paper, we shall often compare our results with what is known in the random utility model (RUM).\footnote{\linespread{1.1} \selectfont Random utility models are studied in \cite{luce1959individual}, \cite{Block1960}, \cite{mcfadden1973conditional}, \cite{falmagne1978representation}, \cite{mcfadden1990stochastic}, \cite{kitamura2018nonparametric}, among others.}

The starting point in these models is a data set with repeated cross-sections of the demand distribution at different prices. To keep our discussion focused on the essentials, we shall assume a discrete cross-sectional data set.  With minor modifications, our results carry through to the case where the demand distributions are not necessarily discrete; this is considered in detail in the Appendix.  Given the discreteness assumption, the data set, consisting of $T$ observations, can be written as
\begin{equation}
\mathcal{D}=\big\{E^t, \overline{\mathbf{p}}^{t}\big\}_{t\in T}.
\end{equation}
At each observation, $t$, the analyst observes the price index $\overline{\mathbf{p}}^{t}$ as well as $N$ expenditure allocation decisions over $K$ goods, which form the set $E^t$.  It is usual in this model to assume that, for any $\mathbf{e}$ and $\mathbf{e}'$ in $E^t$, $\sum_{k=1}^K e_k=\sum_{k=1}^K e'_k=m^t$.  In other words, the total expenditure is the same for different elements in $E^t$ and their implied demand bundles $\mathbf{x}(\mathbf{e},\overline{\mathbf{p}}^t)$ and $\mathbf{x}(\mathbf{e}',\overline{\mathbf{p}}^t)$ are both bundles in the budget set $\mathscr{B}(\overline{\mathbf{p}}^t,m^t)$.

We could think of $E^t$ {\em either} as the allocation decisions of a population with $N$ consumers {\em or} as $N$ allocation decisions made by the same consumer whose demand behavior is stochastic.  We shall refer to a data set of the form $\cal D$ as a \textbf{cross-sectional data set}.

A sorting function at $t$, $\sigma^t$, is a one-to-one map from the set $\{1,2,\ldots, N\}$ to $E^t$; in other words, $\sigma^t$ attaches an index to each expenditure allocation in $E^t$. Note that $\sigma^t(i)$ is an element of $E^t$ that corresponds to $\mathbf{e}^{i, t}$. We say that a cross-sectional data set $\cal D$ is \textbf{RUM-rationalizable} (or admits a RUM-rationalization) if there are sorting functions $\{\sigma^t\}_{t\in T}$ such that
\begin{equation}\label{sortE}
\mathcal{E}(\{\sigma^t\}_{t\in T})=\big\{\big(\sigma^t(i)\big)_{i\in N}, \overline{\mathbf{p}}^{t}\big\}_{t\in T}\mbox{,}
\end{equation}
(which has the same form as (\ref{ourdata set})) is rationalizable with heterogeneous preferences.  In other words, $\cal D$ is RUM-rationalizable if there is a way of sorting the expenditure allocations in $E^t$ so that the entire data set could be understood as arising from $N$ utility-maximizing consumers, all of whom face prices $\overline{\mathbf{p}}^t$ at each observation $t$, but have possibly heterogeneous preferences. We can compare this notion of rationalization with that under the \textbf{random price model} (RPM).  The cross-sectional data set $\cal D$ is  \textbf{RPM-rationalizable} if there are sorting functions $\{\sigma^t\}_{t\in T}$ such that $\mathcal{E}(\{\sigma^t\}_{t\in T})$ (as defined by (\ref{sortE})) is rationalizable by heterogeneous prices, subject to some consistency condition on the heterogeneous prices.

\section{Price Heterogeneity with Consistent Expectations}\label{result1}

In this section, we show that price heterogeneity, even when it is required to be consistent in expectation, is powerful enough to explain almost any data set. Our first result states this claim formally.


\begin{proposition}\label{One} Let $\mathcal{W}=\{W^t_k\}_{k\in K, t\in T}$ be a collection of aggregator functions and $\mathcal{E}=\big\{\big(\mathbf{e}^{i, t}\big)_{i\in N}, \overline{\mathbf{p}}^{t}\big\}_{t\in T}$ be a data set, with $e^{i, t}_k>0$ for all $(i,t)\in N\times T$ and $k=1,2$. Then there are heterogeneous prices ${\cal P}=\{\mathbf{p}^{i,t}\}_{(i,t)\in N\times T}$, which are consistent in expectation w.r.t. $\mathcal{W}$ and satisfy $p^{i,t}_k=\overline{p}_k^t$ for all $k>2$, that rationalize $\mathcal{E}$.
\end{proposition}

In empirical studies of demand, heterogeneity in preferences is often employed as a way of capturing heterogeneity in observed behavior.  This proposition highlights the possibility of an alternative way to explain demand heterogeneity.  It states that if there are two goods for which demand is strictly positive at all observations and for all agents, then price heterogeneity in {\em just those two goods} can rationalize any pattern of expenditure, even if consistency in expectation is imposed.  In the Errors interpretation of heterogeneous prices, the heterogeneity in demand arises from heterogeneity in prices that the analyst could not observe because of imperfections in the data collected.  In the Misperceptions interpretation of heterogeneous prices, the heterogeneity in demand arises from mistakes in consumers' decision-making. Either way, heterogeneous prices could explain heterogeneous demand behavior.  Of course, whether such an explanation is a plausible one would depend on the specific case being considered -- we are not suggesting that it is always and everywhere a better explanation, only that it is a potentially powerful tool for explaining demand heterogeneity.

In the context of a cross-sectional data set ${\cal D}=\big\{E^t, \overline{\mathbf{p}}^{t}\big\}_{t\in T}$, it is well-known that RUM imposes observable restrictions on $\cal D$ (see the example in Section \ref{stable}).  On the other hand, Proposition \ref{One} guarantees that {\em any} data set $\cal D$ with the property that $e^t_k>0$ for $k=1,2$ and all $\mathbf{e}^t\in E^t$ is RPM-rationalizable by heterogeneous prices with consistent price expectations.\footnote{\linespread{1.1} \selectfont Indeed, with {\em any} set of sorting functions, $\{\sigma^t\}_{t\in T}$, Proposition \ref{One} guarantees that $\mathcal{E}(\{\sigma^t\}_{t\in T})=\big\{\big(\sigma^t(i)\big)_{i\in N}, \overline{\mathbf{p}}^{t}\big\}_{t\in T}$ is rationalizable with heterogeneous prices.}  This is good news if the objective is to have a fully flexible model that could capture any set of observed demand behavior on $\mathcal{D}$.  Of course, any explanation of the data with such a method will have implications on the unobserved heterogeneity in prices, and it would then be a matter of (context-dependent) judgment whether the unobserved heterogeneity required to explain the data is, in fact, reasonable.  Another approach is to impose stronger restrictions on the types of price heterogeneity permitted, which will temper the explanatory power of the heterogeneous-prices model; this approach is explored in Section \ref{stable}.

\subsection{Proof of Proposition \ref{One} and related results}  \label{sec:afri}

The proof of Proposition \ref{One}, as well as other results in this paper, relies on Afriat's Theorem (\cite{afriat1967construction}, \cite{varian1982nonparametric}).  The theorem considers a data set with $T$ observations, where observation $t$ consists of a bundle $\mathbf{x}^t\in\mathbb{R}^K_+$ purchased when the prices are $\mathbf{p}^t\in\mathbb{R}^K_{++}$. Thus the data set may be written as $\mathcal{O}=\left\{(\mathbf{x}^t, \mathbf{p}^t)\right\}_{t\in T}$. A strictly increasing and continuous utility function $u:\mathbb{R}^K_+\to\mathbb{R}$ is said to rationalize $\mathcal{O}$ if, for each $t\in T$,
$$u(\mathbf{x}^t)\geq u(\mathbf{x})\:\mbox{ for all $\mathbf{x}\in\mathscr{B}(\mathbf{p}^t,\mathbf{p}^t\cdot\mathbf{x}^t)$}.$$
Afriat's Theorem characterizes rationalizable data sets through a property called the \textbf{generalized axiom of revealed preference}, or \textbf{GARP} for short.\footnote{\linespread{1.1} \selectfont To be precise, $\mathcal{O}$ satisfies GARP if it can be rationalized by a locally nonsatiated preference. Conversely, when $\mathcal{O}$ obeys GARP, then it is rationalizable by a strictly increasing, continuous, and concave utility function.}

GARP is basically a type of no-cycling condition on $\mathcal{X}=\{\mathbf{x}^{t}\}_{t\in T}$, the set of observed demand bundles. Let $\mathbf{x}^t$, $\mathbf{x}^s$ be two elements in $\mathcal{X}$. We say $\mathbf{x}^t$ is \textbf{directly revealed preferred to} $\mathbf{x}^s$, denoted by $\mathbf{x}^t\succsim_D \mathbf{x}^s$, if $\mathbf{p}^t\cdot \mathbf{x}^t\ge \mathbf{p}^t\cdot \mathbf{x}^s$. Similarly, $\mathbf{x}^t$ is \textbf{directly revealed strictly preferred to} $\mathbf{x}^s$, denoted by $\mathbf{x}^t\succ_D \mathbf{x}^s$, if $\mathbf{p}^t\cdot \mathbf{x}^t>\mathbf{p}^t\cdot \mathbf{x}^s$. We say $\mathbf{x}^t$ is \textbf{revealed preferred to} $\mathbf{x}^s$, denoted by $\mathbf{x}^t\succsim_R \mathbf{x}^s$, if there is a sequence $\{\mathbf{x}^{t_l}\}^L_{l=1}$ such that $\mathbf{x}^{t_1}=\mathbf{x}^t$, $\mathbf{x}^{t_L}=\mathbf{x}^s$, and $\mathbf{x}^{t_l}\succsim_D \mathbf{x}^{t_{l+1}}$ for each $l<L$. GARP requires that for any $s, t\in T$, $\mathbf{x}^t\succsim_R \mathbf{x}^s$ implies $\mathbf{x}^s\not\succ_D \mathbf{x}^t$.

In essence, the proof of Proposition \ref{One} consists of constructing heterogeneous prices for goods 1 and 2 that are consistent in expectation and have the property that the collection of observations
\begin{equation}
\mathcal{O}^*=\left\{\left(\mathbf{x}(\mathbf{e}^{i,t}, \mathbf{p}^{i,t}),\mathbf{p}^{i,t}\right)\right\}_{(i,t)\in N\times T}
\end{equation}
obeys GARP.  This is a notional data set with $N\times T$ observations, where each observation consists of a bundle and a price vector, and is precisely of the type where one may apply Afriat's Theorem.  The theorem guarantees that there is a utility function that rationalizes this $\mathcal{O}^*$ (equivalently, rationalizes $\mathcal{E}$ with heterogeneous prices) if $\mathcal{O}^*$ satisfies GARP. In our proof, we choose the idiosyncratic prices $p^{i,t}_k$, for $k=1,2$, carefully so that the direct revealed preference relation $\succsim_D$ on $\mathcal{X}^*=\{\mathbf{x}(\mathbf{e}^{i,t}, \mathbf{p}^{i,t})\}_{(i,t)\in N\times T}$ has the following property:
\begin{equation}
\text{if }\, \mathbf{x}(\mathbf{e}^{i, t}, \mathbf{p}^{i, t})\succsim_{D} \mathbf{x}(\mathbf{e}^{j, s}, \mathbf{p}^{j, s}),\text{ then either }i>j\text{ or }i=j\text{ and }t\geq s.
\end{equation}
In other words, if $\mathbf{x}(\mathbf{e}^{i, t}, \mathbf{p}^{i, t})$ is directly revealed preferred to $\mathbf{x}(\mathbf{e}^{j, s},\mathbf{p}^{j, s})$, then $(i,t)$ dominates $(j,s)$ in the lexicographic order on $N\times T$.\footnote{\linespread{1.1} \selectfont The lexicographic order ranks $(i, t)$ higher than $(j, s)$ if either $i>j$ or $i=j$ and $t>s$.} This in turn guarantees that if $\mathbf{x}(\mathbf{e}^{i, t}, \mathbf{p}^{i, t})\succsim_{R} \mathbf{x}(\mathbf{e}^{j, s}, \mathbf{p}^{j, s})$, then either $i>j$ or $i= j$ and $t\geq s$.  Thus, it is impossible for
$\mathbf{x}(\mathbf{e}^{i, t}, \mathbf{p}^{i, t})\succsim_{R} \mathbf{x}(\mathbf{e}^{j, s}, \mathbf{p}^{j, s})$ and for
$\mathbf{x}(\mathbf{e}^{j, s}, \mathbf{p}^{j, s})\succ_D \mathbf{x}(\mathbf{e}^{i, t}, \mathbf{p}^{i, t})$, which means that GARP holds.  The details of the proof can be found in the Appendix.
\medskip

\noindent\textbf{Relationship to \cite{varian1988revealed}.}\,  \cite{varian1988revealed} shows that GARP becomes a vacuous condition in certain circumstances when prices or demand are only partially observed. To be specific, suppose that at each observation $t$, the prices of all $K$ goods, $\mathbf{p}^t$, are observed, but only the consumer's demand for goods $2,3,\ldots, K$, which we denote by $\mathbf{x}^t_{-1}=(x^t_2, \ldots, x^t_{K})$, are observed.  In this case, there would always exist $\hat x^t_1\in\mathbb{R}_+$ (a hypothetical demand for good 1 at observation $t$) such that $\{(\mathbf{x}^t, \mathbf{p}^t)\}_{t\in T}$ obeys GARP, where $\mathbf{x}^t=(\hat x^t_1,\mathbf{x}_{-1}^t)$. In other words, when the demand of one good is missing from the data, rationalizability can never be rejected. The proof provided by Varian involves picking a sequence $x^t_1$ (for $t=1,2,\ldots$) in such a way that it is impossible for $\mathbf{x}^t$ to be revealed preferred to $\mathbf{x}^{t'}$ if $t<t'$.

While not explicitly considered in \cite{varian1988revealed}, it is clear that Varian's result has a variation of the following form (which is also clear from our proof of Proposition \ref{One}): {\em suppose each observation $t$ consists of the prices and demand for goods $2,3,\ldots, K$ (denoted by $\mathbf{p}^t_{-1}$ and $\mathbf{x}^t_{-1}$ respectively) as well as the expenditure on good $1$, $e^t_1$, where $e^t_1>0$ for all $t$; then there is $\hat p^t_1$ and $\hat x^t_1$ such that $\hat p^t_1\hat x^t_1=e^t_1$ and, with $\mathbf{x}^t=(\hat x^t_1,\mathbf{x}^t_{-1})$ and $\mathbf{p}^t=(\hat p^t_1,\mathbf{p}^t_{-1})$, the data set $\{(\mathbf{x}^t, \mathbf{p}^t)\}_{t\in T}$ obeys GARP (and is thus rationalizable).}

Compared to this result, the setup of Proposition \ref{One} is more complicated because $\mathcal{E}=\big\{\big(\mathbf{e}^{i, t}\big)_{i\in N}, \overline{\mathbf{p}}^{t}\big\}_{t\in T}$ collects observations from multiple agents and the average price of each good (across all agents) is observed.  Thus, while for a {\em given} agent $i$ we are free to `break up' the expenditure on good 1, $e^{i,t}_1$, in any way between the price and demand for good 1, a higher idiosyncratic price of good 1 for one agent has to be balanced by a lower price of good 1 for another agent.  With this added consistency requirement, the rationalizability of $\cal E$ can only be guaranteed if there is price heterogeneity in at least two goods.  In the Appendix, we provide an example of a data set $\cal E$ that cannot be rationalized with idiosyncratic prices in only one good, even if agents are permitted to have distinct preferences.
\medskip

\noindent\textbf{Relationship to \cite{brown1996testable}.} \cite{brown1996testable} show that the Walrasian model of an exchange economy is refutable when the economy's total endowment, the equilibrium prices, and the equilibrium income distribution are observable. Their result could be re-formulated (as we do below) as a result on the dis-aggregation of demand.  

The data set consists of $T$ observations with each observation $t$ consisting of the following:\, {\em aggregate} expenditure (across all consumers) on each good $k$, $E^t_k>0$;  the aggregate expenditure of each consumer, $m^{i,t}>0$; and the prevailing price vector $\overline{\mathbf{p}}^t\gg 0$. Formally, 
$${\cal E}^{**}=\left\{(E^{t}_k)_{k\in K},(m^{i,t})_{i\in N},\overline{\mathbf{p}}^t\right\}_{t\in T}.$$ 
(For consistency, we require $\sum_{k\in K} E_k^t=\sum_{i\in N}m^{i,t}$ at each $t$; in other words, total expenditure on all goods equals total expenditure of all consumers, at each $t$.)  We refer to $(e^{i,t}_k)_{(i,k,t)\in N\times K\times T}\geq 0$ as a disaggregation of $(E^{t}_k)_{(k,t)\in K\times T}$ if $\sum_{i\in N}e^{i,t}_k=E^t_k$ for each $(t,k)$ and $\sum_{k\in K}e^{i,t}_k=m^{i,t}$ for each $(i,t)$. We say that ${\cal E}^{**}$ admits a {\bf disaggregation with heterogeneous preferences} if there is a disaggregation $(e^{i,t}_k)_{(i,k,t)\in N\times K\times T}$ such that ${\cal E}=\{(\mathbf{e}^{i,t})_{i\in N},\overline{\mathbf{p}}^t\}_{t\in T}$ is rationalizable with heterogeneous preferences; similarly, ${\cal E}^{**}$ admits a {\bf disaggregation with heterogeneous prices} if there is a disaggregation such that  ${\cal E}$ is rationalizable with heterogeneous prices.   

One may be tempted to think that {\em every} ${\cal E}^{**}$ admits a disaggregation with heterogeneous preferences, but \cite{brown1996testable} show that there exists ${\cal E}^{**}$ that does not admit such a disaggregation; they also set out a procedure that could test if such a disaggregation exists. This result is sensitive to the assumption of uniform pricing across consumers, and price heterogeneity on any two goods will guarantee that `anything goes.'   

\begin{corollary} \label{BM-decom}
Let $\mathcal{W}=\{W^t_k\}_{k\in K, t\in T}$ be a collection of aggregator functions. For every data set ${\cal E}^{**}=\{(E^{t}_k)_{k\in K},(m^{i,t})_{i\in N},\overline{\mathbf{p}}^t\}_{t\in T}$ there is a disaggregation $(e^{i,t}_k)_{(i,k,t)\in N\times K\times T}$ with the following property:\, for any distinct $k',k''\in K$, there are heterogeneous prices ${\cal P}=\{\mathbf{p}^{i,t}\}_{(i,t)\in N\times T}$ that are consistent in expectation w.r.t. $\mathcal{W}$, satisfy $p^{i,t}_k=\overline{p}_k^t$ for all $k\neq k',\, k''$, and rationalize ${\cal E}=\{(\mathbf{e}^{i,t})_{i\in N},\overline{\mathbf{p}}^t\}_{t\in T}$. 
\end{corollary}

This corollary follows immediately from Proposition \ref{One}, by simply choosing the disaggregation $e^{i,t}_k=[m^{i,t}/(\sum_{i\in N}m^{i,t})]E^{t}_k>0$. In the next section, we show that stronger restrictions on price heterogeneity will diminish the explanatory power of price heterogeneity and allow us to restore the Brown-Matzkin result.

\section{Stable Price Heterogeneity and Preference Heterogeneity}\label{stable}

In the previous section, we analyzed price heterogeneity with consistent expectations and found that it gives rise to a very flexible model that imposes no restrictions on observed demand.  In this section, we explore the implications of stable price heterogeneity. In Section 4.1, we explain why stable price heterogeneity does not enlarge the observable implications of a model with heterogeneous preferences and homogeneous prices.  Section 4.2 contains the main result of the paper: we show that stable price heterogeneity has precisely the same explanatory power as heterogeneous preferences.  

\subsection{Preference heterogeneity when prices are heterogeneous}

In the result below, we ask what happens when there is stable price heterogeneity (i.e., $p^{i, t}_k=\lambda_{i, k}\,\overline{p}^{t}_k$) in a market where consumers have heterogeneous preferences.

\begin{proposition} \label{prop-indistinct}
For any stable price distribution ${\cal P}=\{\mathbf{p}^{i,t}\}_{(i,t)\in N\times T}$, the data set $\mathcal{E}=\{(\mathbf{e}^{i, t})_{i\in N}, \overline{\mathbf{p}}^{t}\}_{t\in T}$ is rationalizable with heterogeneous preferences if and only if, for each $i\in N$, $\{(\mathbf{e}^{i, t}, \mathbf{p}^{i, t})\}_{t\in T}$ is rationalizable; i.e., there is a strictly increasing and continuous utility function $u^i:\mathbb{R}^K_+\to\mathbb{R}$ such that
for each $t\in T$,
\[u^i(\mathbf{x}(\mathbf{e}^{i,t}, \mathbf{p}^{i, t})) \ge u^i(\mathbf{x}) \text{ for any }\mathbf{x}\in \mathscr{B}\left(\mathbf{p}^{i, t}, m^{i,t}\right).\]
\end{proposition}

The proof of this proposition uses Afriat's Theorem and can be found in the Appendix. To interpret this result, we first consider the case where prices vary across consumers ({\em in fact}) with $\overline{\mathbf{p}}^{t}$ being the average price vector at observation $t$.  Then Proposition \ref{prop-indistinct} tells us that, provided each agent's demand is rationalizable and price heterogeneity is stable, the average price $\overline{\mathbf{p}}^{t}$ is a good substitute for the true price $\mathbf{p}^{i, t}$, in the sense that it does not destroy the rationalizability of the data.  In the second interpretation, where $\mathbf{p}^{i, t}$ is agent $i$'s perceived price, this proposition tells us that stable price misperception is undetectable: the agent's behavior is indistinguishable from someone who perceives prices correctly and is maximizing utility.

Our result supports this practice of focusing on preference heterogeneity yet ignoring price heterogeneity in empirical work, even when price heterogeneity is prevalent. 

\medskip

\noindent\textbf{Cross-Sectional Data Environment.} Proposition  \ref{prop-indistinct} has the following immediate corollary for cross-sectional data sets: {\em for any stable price distribution ${\cal P}=\{\mathbf{p}^{i,t}\}_{(i,t)\in N\times T}$, the cross-sectional data set ${\cal D}=\{E^t,\overline{\mathbf{p}}^t\}_{t\in T}$ is RUM-rationalizable if and only if there are sorting functions $\{\sigma^t\}_{t\in T}$ and strictly increasing and continuous utility functions $u^i:\mathbb{R}^K_+\to\mathbb{R}$ such that for each $i\in N$ and $t\in T$,
\begin{equation}
u^i(\mathbf{x}(\sigma^t(i), \mathbf{p}^{i, t})) \ge u^i(\mathbf{x}) \text{ for any }\mathbf{x}\in \mathscr{B}\left(\mathbf{p}^{i, t}, m^t\right).
\end{equation}}  In other words, RUM-rationalizability is not destroyed by the insertion of stable price heterogeneity:  if ${\cal D}$ is a cross-sectional data set collected from a population of consumers who maximize (heterogeneous) increasing and continuous utility functions under prices with stable heterogeneity, then $\cal D$ will be RUM-rationalizable. It follows that the observable implications of the random utility model (such as that depicted in Figure 1) remain precisely the same even if there is stable price heterogeneity among agents in the population.\medskip

\begin{figure} \label{fig-rum}

\centering
\begin{tikzpicture}[x=0.65cm, y=0.65cm][domain=0:1, range=0:1, scale=3/4, thick]
\usetikzlibrary{intersections}

\draw[very thick][->] (10,0)--(10,8) node[rotate=90, anchor=south east]{};
\draw[very thick][->] (10,0)--(19,0) node[anchor=north east]{};

\draw[](10, 7)--(14, 0);

\draw[](10, 3)--(18, 0);

\filldraw[thick] (13.7, 0.55) circle (2pt);

\filldraw[thick] (10.8, 2.7) circle (2pt);
\filldraw[thick] (11.9, 2.3) circle (2pt);

\coordinate[label=left:${B_1}$] (wf) at (11.5, 6.6);

\coordinate[label=below:${B_2}$] (wf) at (17.6, 1.2);

\coordinate[label=left:${x_2}$] (wf) at (10, 7.5);

\coordinate[label=below:${x_1}$] (wf) at (18.5, 0);

\coordinate[label=above:${\footnotesize{\mathbf{a}'}}$] (wf) at (14.42, 0.325);

\filldraw[thick] (11, 5.25) circle (2pt);
\coordinate[label=above:${\footnotesize{\mathbf{a}}}$] (wf) at (10.8, 4.3);

\coordinate[label=above:${\footnotesize{\mathbf{b}'}}$] (wf) at (11.5, 1.3);
\coordinate[label=above:${\footnotesize{\mathbf{b}}}$] (wf) at (10.6, 1.6);


\draw[very thick][->] (20,0)--(20,8) node[rotate=90, anchor=south east]{};
\draw[very thick][->] (20,0)--(29,0) node[anchor=north east]{};

\draw[](20, 7)--(24, 0);

\draw[](20, 3)--(28, 0);

\filldraw[thick] (23.7, 0.55) circle (2pt);

\filldraw[thick] (20.8, 2.7) circle (2pt);
\filldraw[thick] (26, 0.75) circle (2pt);

\coordinate[label=left:${x_2}$] (wf) at (20, 7.5);

\coordinate[label=below:${x_1}$] (wf) at (28.5, 0);

\coordinate[label=above:${\footnotesize{\mathbf{a}'}}$] (wf) at (24.42, 0.325);

\filldraw[thick] (21, 5.25) circle (2pt);
\coordinate[label=above:${\footnotesize{\mathbf{a}}}$] (wf) at (20.8, 4.3);

\coordinate[label=above:${\footnotesize{\mathbf{b}'}}$] (wf) at (26.5, 0.8);
\coordinate[label=above:${\footnotesize{\mathbf{b}}}$] (wf) at (20.6, 1.6);

\end{tikzpicture}

\caption{Observable restrictions in the Random Utility Model}
\vspace{-.1in}
\caption*{\linespread{1.1}\selectfont\footnotesize The cross-sectional data set depicted on the right, with $E^1=\{\mathbf{a},\mathbf{a}'\}$ and $E^2=\{\mathbf{b},\mathbf{b}'\}$ is RUM-rationalizable since one could pair up $\mathbf{a}$ with $\mathbf{b}$ and $\mathbf{a}'$ with $\mathbf{b}'$ and each of the agent-level data sets obeys GARP.  On the other hand, the data set on the left is not RUM-rationalizable because it is impossible to sort the elements of $E^1$ and $E^2$ in such a way that both agent-level data sets obey GARP.}
\end{figure}

\noindent\textbf{Relationship to \cite{brown1996testable} (continued).}  In Corollary \ref{BM-decom}, we show that {\em every} data set ${\cal E}^{**}=\{(E^{t}_k)_{k\in K},(m^{i,t})_{i\in N},\overline{\mathbf{p}}^t\}_{t\in T}$ admits a decomposition with heterogeneous prices which are consistent in expectation.  This `anything goes' result no longer holds if we confine ourselves to stable price heterogeneity.  Indeed, it follows immediately from Proposition \ref{prop-indistinct} that, for any stable price distribution ${\cal P}=\{\mathbf{p}^{i,t}\}_{(i,t)\in N\times T}$, the data set ${\cal E}^{**}$ admits a disaggregation $\mathcal{E}=\{(\mathbf{e}^{i, t})_{i\in N}, \overline{\mathbf{p}}^{t}\}_{t\in T}$ that is rationalizable with heterogeneous preferences if and only if, for each $i\in N$, $\{(\mathbf{e}^{i, t}, \mathbf{p}^{i, t})\}_{t\in T}$ is rationalizable (by a strictly increasing and continuous utility function). In particular, the observable restrictions on $\cal E^{**}$ imposed by heterogeneous preferences and homogeneous prices are {\em exactly the same} as the observable restrictions imposed by heterogeneous prices and stable price heterogeneity.  Thus the Brown-Matzkin result is robust to the presence of stable price heterogeneity.

\subsection{Stable Price Heterogeneity and Equivalence Scale Utilities}

Our objective in this section is to establish a threefold equivalence.  Our first equivalence is to show that if $\mathcal{E}=\{(\mathbf{e}^{i, t})_{i\in N}, \overline{\mathbf{p}}^{t}\}_{t\in T}$ is rationalizable with heterogeneous preferences, then it suffices to choose these preferences from a semiparametric family of utility functions where (in a specific sense) any two utility functions in the family are related by a linear transformation.  The second equivalence is that $\cal E$ is rationalizable with heterogeneous preferences if and only if it is rationalizable with heterogeneous prices with a stable distribution.

Let $R$ be a nonempty subset of $K$. For a utility function $U:\mathbb{R}^{K}_{+}\to\mathbb{R}$ and $\beta\in\mathbb{R}_{++}$, let
\begin{equation*}
U^\beta_R(\mathbf{x})\equiv U(\beta\,\mathbf{x}_R, \mathbf{x}_{-R})\:\:\mbox{ for any $\mathbf{x}\in \mathbb{R}^{K}_{+}$.}
\end{equation*}
We refer to $U^\beta_R$ as a {\em $R$-scale transformation} of $U$.  A collection $\mathscr{U}_R$ of utility functions forms an \textbf{$R$-scale family} if there is a strictly increasing and concave utility function $U$ such that $\mathscr{U}_R=\{U^\beta_R: \beta\in\mathbb{R}_{++}\}$.  Scale transformations first appeared in the econometric literature on demand estimation, where they are used to capture (among other things) the effects of household composition on demand (see, for example, \cite{barten1964consumer}, \cite{prais1971analysis}, and \cite{muellbauer1980estimation}).  They have also been used in the study of demand aggregation, where they help to obtain properties on {\em market} demand which are not necessarily present at the agent-level demand correspondence (see \cite{mas1977some}, \cite{dierker1984price}, \cite{grandmont1987distributions, grandmont1992transformations}, and \cite{quah1997law}).

A basic reason for the use of scale transformations is that there is a convenient relationship between the demand generated by $U$ and that generated by $U^\beta_R$.  Let $\mathbf{x}(\mathbf{p}, m)$ and $\mathbf{x}^\beta(\mathbf{p}, m)$ be their demand sets at price $\mathbf{p}$ and income/expenditure $m$, i.e., $\mathbf{x}(\mathbf{p}, m)=\argmax\{U(\mathbf{x})|\mathbf{x}\in \mathscr{B}(\mathbf{p}, m)\}$ and $\mathbf{x}^\beta(\mathbf{p}, m)=\argmax\{U^\beta_R(\mathbf{x})|\mathbf{x}\in \mathscr{B}(\mathbf{p}, m)\}$. Then it is straightforward to check that $\mathbf{x}$ and $\mathbf{x}^\beta$ are related in the following way:
\[\mathbf{x}^\beta(\mathbf{p}, m)=\Big(\frac{\mathbf{x}_R(\mathbf{p}, \beta\, m)}{\beta}, \mathbf{x}_{-R}(\mathbf{p}, \beta\, m)\Big).\]
Note that in the case where $R=K$, we have $\mathbf{x}^\beta(\mathbf{p}, m)=\mathbf{x}(\mathbf{p}, \beta\, m)\big{/}\beta$.\footnote{\linespread{1.1} \selectfont Moreover, when $R=K$, the Independence of a Base Utility (IB) property in \cite{lewbel1989household} (also called \textit{Equivalence Scale Exactness}) is satisfied. The IB property is commonly assumed in empirical demand work to justify the use of the cost-of-living indices.}

We have explained in Section \ref{model} what it means for a data set $\mathcal{E}=\{(\mathbf{e}^{i, t})_{i\in N}, \overline{\mathbf{p}}^{t}\}_{t\in T}$ to be  rationalizable with heterogeneous preferences.  We say that $\cal E$ is rationalizable with heterogeneous preferences \textbf{drawn from an $R$-scale family} if the preferences needed to explain $\cal E$ can all be induced by utility functions drawn from an $R$-scale family; more formally, there are real numbers $\beta_1, \ldots, \beta_N\in \mathbb{R}_{++}$ and a strictly increasing and concave utility function $U$ such that for each $(i, t)\in N\times T$,
\begin{equation}
U^{\beta_i}_R(\mathbf{x}(\mathbf{e}^{i,t}, \overline{\mathbf{p}}^{t})) \ge U^{\beta_i}_R(\mathbf{x}) \text{ for any }\mathbf{x}\in \mathscr{B}\left(\overline{\mathbf{p}}^{t}, m^{i,t}\right).
\end{equation}

The next result states the threefold equivalence between three distinct notions of rationalizability.

\begin{theorem} \label{thm-four}
Let $\mathcal{E}=\{(\mathbf{e}^{i, t})_{i\in N}, \overline{\mathbf{p}}^{t}\}_{t\in T}$ be a data set and suppose that for some $ R\subseteq K$, we have $\mathbf{e}^{i, t}_R>0$ for each $(i, t)\in N\times T$.  Then the following are equivalent.\vspace{-0.05in}
\begin{itemize}
\item[{\em (i)}] $\mathcal{E}$ is rationalizable with heterogeneous preferences.\vspace{-0.05in}
\item[{\em (ii)}] $\cal E$ is rationalizable with heterogeneous preferences drawn from an $R$-scale family.\vspace{-0.05in}
\item[{\em (iii)}] Let $W$ be any linear homogeneous aggregator function. Then there are real numbers $\lambda_1, \ldots, \lambda_N>0$ such that the stable price distribution ${\cal P}=\{\mathbf{p}^{i,t}\}_{(i,t)\in N\times T}$ with $\mathbf{p}^{i, t}=(\lambda_i\, \overline{\mathbf{p}}^t_R, \overline{\mathbf{p}}^t_{-R})$ is consistent in expectation w.r.t. $W$ and rationalizes $\mathcal{E}$.
\end{itemize}
{\em Note:\, Since our maintained assumption is that $\cal E$ satisfies $\sum_{k\in K}e^{i, t}_k >0$ for all $(i,t)$, we could always choose $R=K$,  but the theorem allows for $R$ to be strictly smaller.  If (say) good 1 satisfies $e^{i,t}_1>0$ for all $(i,t)$, then we could choose $R=\{1\}$.  Statement (ii) then refers to rationalizability with preference heterogeneity obtained by scaling good 1, while statement (iii) refers to rationalizability with heterogeneity in just the price of good 1; each of these statements is equivalent to rationalizability with heterogeneous preferences.}
\end{theorem}

In the literature on demand aggregation, it is fairly common to assume that all agents in a market (or some segment of the market) have preferences that belong to an $R$-scale family (see, for example, \cite{grandmont1987distributions, grandmont1992transformations} and \cite{quah1997law}).  The equivalence of statements (i) and (ii) in this theorem tells us that this modeling practice is (in essence) without loss of generality since any data set $\cal E$ that is rationalizable with heterogeneous preferences can also be rationalized by preferences drawn from an $R$-scale family.  Of course, in that literature, further conditions are imposed on the distribution of the scales (that is, the distribution of $\beta$ in our notation) and it is these further distributional assumptions that drive the strong conclusions on market demand obtained in that literature.\footnote{\linespread{1.1} \selectfont For example, sufficient dispersion in the distribution of $\beta$ leads to the approximate linearity of the market demand function with respect to total expenditure/income (see \cite{grandmont1992transformations} and \cite{quah1997law}).}

Recall that Proposition \ref{One} tells us that {\em any} data set $\cal E$ can be explained by heterogeneous prices, even when this heterogeneity is required to satisfy a consistency condition.  The equivalence of (i) and (iii) in Theorem \ref{thm-four} tells us that the imposition of stable heterogeneity on prices disciplines the implications of heterogeneous prices; in fact, the explanatory power of such a model (in the sense of the data sets that it could explain) coincides precisely with that of a model with heterogeneous preferences. In the case where there is a good -- say, good 1 -- where the expenditure is strictly positive for all agents and at all observations, then (stable) price heterogeneity for that good alone is sufficient to capture the effects of random preferences.  Lastly, note that the combination of Proposition \ref{prop-indistinct} and Theorem \ref{thm-four} means that even when preference heterogeneity and stable price heterogeneity are {\em both} present in a given environment, the data it generates is equivalent to data from an environment where {\em only} preference heterogeneity or price heterogeneity is present.

Note that our result supports this common practice of using the equivalence scale utilities in empirical work.

The next result is an obvious corollary of Theorem \ref{thm-four} obtained by transposing the theorem to the setting of a cross-sectional data set.  The punchline is that {\em RUM-rationalizability is equivalent to RPM-rationalizability with stable heterogeneous prices}. In the Appendix, we provide a more general formulation of this equivalence that allows the cross-sectional data set to be non-discrete (as in \cite{mcfadden1990stochastic} or \cite{kitamura2018nonparametric}).

\begin{corollary}\label{thm-five} Let $\mathcal{D}=\big\{E^t, \overline{\mathbf{p}}^{t}\big\}_{t\in T}$ be a cross-sectional data set and suppose that for some $ R\subseteq K$, we have $\mathbf{e}^{i, t}_R>0$ for each $(i, t)\in N\times T$.  Then, the following are equivalent.\vspace{-0.05in}
\begin{itemize}
\item[{\em (i)}] $\mathcal{D}$ is RUM-rationalizable.\vspace{-0.05in}
\item[{\em (ii)}] $\cal D$ is RUM-rationalizable with heterogeneous preferences drawn from an $R$-scale family.\vspace{-0.05in}
\item[{\em (iii)}] Let $W$ be any linear homogeneous aggregator function. Then $\cal D$ is RPM-rationalizable with a price distribution $\cal P$ that is stable and consistent in expectation w.r.t. $W$, with heterogeneous prices only for the goods in $R$.
\end{itemize}
\end{corollary}

\noindent {\bf Proof sketch of Theorem \ref{thm-four}.}\, The detailed proof is found in the Appendix, but the following quick sketch may be helpful.

Obviously, (ii) implies (i).  To show that (iii) implies (ii), let $R\subseteq K$ with $\text{e}^{i, t}_R>0$ for each $(i, t)\in N\times T$. Then there are real numbers $\lambda_1, \ldots, \lambda_N>0$ and an increasing and continuous utility function $U$ such that, with $\mathbf{p}^{i, t}=(\lambda_i\, \overline{\mathbf{p}}^t_R, \overline{\mathbf{p}}^t_{-R})$,
\begin{equation}
U(\mathbf{x}(\mathbf{e}^{i,t}, \mathbf{p}^{i,t})) \ge U(\mathbf{x}) \text{ for any }\mathbf{x}\in \mathscr{B}\left(\mathbf{p}^{i,t}, m^{i, t}\right)
\end{equation}
for all $(i,t)$.  We show that this implies that
\begin{equation*}
U^{\frac{1}{\lambda_i}}_R(\mathbf{x}(\mathbf{e}^{i,t}, \overline{\mathbf{p}}^t))\ge U^{\frac{1}{\lambda_i}}_R(\tilde{\mathbf{x}})\text{ for any }\tilde{\mathbf{x}}\in\mathscr{B}\left(\overline{\mathbf{p}}^{t}, m^{i, t}\right)
\end{equation*}
and thus $\mathcal{E}$ is rationalizable with heterogeneous preferences drawn from an $R$-scale family. It remains to show that (i) implies (iii).  Suppose $\mathcal{E}$ is rationalizable with heterogeneous preferences. We need to find real numbers $\lambda_1, \ldots, \lambda_N>0$ such that  $$\mathcal{O}^*=\left\{\left(\mathbf{x}(\mathbf{e}^{i,t}, \mathbf{p}^{i,t}),\mathbf{p}^{i,t}\right)\right\}_{(i,t)\in N\times T}\mbox{ where $\mathbf{p}^{i, t}=(\lambda_i\, \overline{\mathbf{p}}^t_R, \overline{\mathbf{p}}^t_{-R})$}$$ 
obeys GARP. Obviously, $\mathcal{O}^*=\bigcup_{i\in N}\mathcal{O}^i$, where $\mathcal{O}^i=\left\{\left(\mathbf{x}(\mathbf{e}^{i,t}, \mathbf{p}^{i,t}),\mathbf{p}^{i,t}\right)\right\}_{t\in T}$. Note that $\mathcal{O}^i$ obeys GARP for any $\lambda_i>0$; this follows from Proposition \ref{prop-indistinct} and the fact that $\mathcal{E}$ is rationalizable with heterogeneous preferences.  We then show that it is possible to choose $\lambda_{i+1}$ sufficiently smaller than $\lambda_i$ so that, for any two consumers $i$ and $j$ with $j>i$, their budget sets are sufficiently distinct so that no bundle chosen by $i$ is revealed preferred to a bundle chosen by $j$.  We could then conclude that $\mathcal{O}^*$ obeys GARP. \hfill \qed

\section{Expenditure-Augmented Utility and Stable Price Heterogeneity} \label{sec-augment}

Let $\mathcal{O}=\left\{(\mathbf{x}^t, \mathbf{p}^t)\right\}_{t\in T}$ be the data set collected from a single consumer, where $\mathbf{x}^t$ is the bundle purchased at the price $\mathbf{p}^t$.  In Section \ref{sec:afri} we recounted Afriat's Theorem, which provides necessary and sufficient conditions for $\cal O$ to be rationalizable, in the sense that there is a utility function such that the chosen bundle $\mathbf{x}^t$ is weakly preferred to alternative bundles in the budget set $\mathscr{B}(\mathbf{p}^t,\mathbf{p}^t\cdot\mathbf{x}^t)$.  Afriat's Theorem addresses the problem of rationalizability in the context of the constrained-optimization model of consumer demand, but there are other ways in which consumer demand can be modeled.

In partial equilibrium settings, where the set of goods being studied is just a subset of all the goods consumed by a consumer, it is common to model demand with a utility function of the quasilinear form; the consumer's choice is then the bundle $\mathbf{x}\in\mathbb{R}^K_+$ that maximizes $\widebar U(\mbx)-\mbp\cdot\mbx$, when $\mbp$ is the price vector for the $K$ goods.  More generally, we can model the consumer's choice as arising from the unconstrained maximization of a utility function where expenditure enters as an argument of the utility function. Formally, the utility of acquiring the bundle $\mbx$ at cost $e\geq 0$ is $V(x,-e)$ where $V:\mathbb{R}^K_+\times \mathbb{R}_{-}\!\!\to\!\mathbb{R}$ is required to be strictly increasing in the last argument (so that increasing expenditure strictly reduces utility). Quasilinear utility-maximization would then correspond to the special case where $V(\mbx,-e)=\bar U(\mbx)-e$.  We refer to $V$ as an {\bf expenditure-augmented utility function} or simply an augmented utility function.

Consumer demand arising from the maximization of such a utility function is studied in \cite{deb2023revealed}.  This study provides necessary and sufficient conditions under which a data set $\cal O$ is augmented utility-rationalizable, in the sense that there is an augmented utility function $V$ for which
$$V(\mbx^t,-\mbp^t\cdot \mbx^t)\geq V(\mbx,-\mbp^t\cdot \mbx)\:\:\mbox{ for all $\mbx\in\mathbb{R}^K_+$.}\footnote{The characterization of those data sets $\cal O$ that are quasilinear-rationalizable, in the sense that there is $\widebar U$ such that
$\bar U(\mbx^t)-\mbp^t\cdot \mbx^t\geq \bar U(\mbx)-\mbp^t\cdot \mbx$ for all $\mbx\in\mathbb{R}^K_+$, can be found in \cite{brown2007nonparametric}.}$$

In this section, we study the implications of price heterogeneity in the augmented utility model (AUM).  We focus on two notions of price heterogeneity, both of which are inspired by the behavioral economics literature. The first notion is price heterogeneity arising from {\bf inattention to prices} among consumers, leading them to optimize according to idiosyncratic prices that differ from the observed price. While we have already introduced this idea in the previous sections, one could argue that the augmented utility model is a particularly convenient setting in which to study this phenomenon:  since this model is not a constrained-optimization model, we can assume that a consumer who has been inattentive to prices and who has chosen a bundle based on a wrong impression of prices will simply {\em purchase the bundle he intended}; in contrast, in a constrained-optimization model, such an assumption cannot be made because a consumer who chooses a bundle based on the wrong prices could violate the budget constraint and thus a reconciliation procedure must be added to the model.    

For example, imagine a consumer $i$ who on a shopping trip to a grocery store chooses a bundle that maximizes an augmented utility function.  However, consumer $i$ is not fully attentive to prices and so chooses the bundle $\mbx$ that maximizes $V(\mbx,-\mbp^i\cdot\mbx)$, where $\mbp^i$ is the consumer's impression of prices, which may differ from the true prices $\mbp$.  We assume that if $\tilde{\mbx}$ is the outcome of this maximization problem, then $\tilde{\mbx}$ is the bundle bought.  The consumer may realize at some point that he has made a mistake and total expenditure is different from what he expected, but since groceries form just a subset of the goods purchased by this consumer, it is plausible to assume that he could adjust for this mistake when purchasing other goods in the future.

The second notion of price heterogeneity we have in mind arises from {\bf reference price dependence}, of the type investigated in \cite{koszegi2006model} and \cite{heidhues2008competition}.  In this case, the consumer observes prices perfectly but has some expectation (or a reference) of what prices should be, and his demand is affected by the extent to which the actual prices he encounters differ from his reference.  Heterogeneous reference prices are then a source of heterogeneity in demand.

\subsection{AU-rationalizability with heterogeneous prices}\label{sec5-result}

We now explain how we can formally incorporate both price inattention and reference price dependence into the augmented utility model.  We assume that a consumer $i$ has in mind a price vector $\mbp^i$, which may be different from the true price vector $\mbp$.   We interpret $\mbp^i$ as either what the consumer thinks are the prevailing prices (which could be wrong because he is inattentive to prices) or as the consumer's reference prices. Let $e=\mbp\cdot \mbx$ and $e^i=\mbp^i\cdot \mbx$.  We assume that the consumer behaves as though he attributes to the bundle $\mbx$ an expenditure level that depends on $e$ and $e^i$; formally, the expenditure associated with the bundle $\mbx$ is $\phi (\mbp\cdot \mbx,\mbp^i\cdot \mbx)$, where $\phi:\mathbb{R}^2_{++}\to\mathbb{R}_+$ has as its arguments $e$ and $e^i$.  We refer to $\phi$ as the {\bf behavioral expenditure function}.  The consumer $i$ chooses $\mbx$ to maximize
$$V(\mbx,-\phi (\mbp\cdot \mbx, \mbp^i\cdot \mbx)),$$
where $V$ is the augmented utility function.

For example, if we are modeling consumers who are inattentive to prices, we let $\phi(e,e^i)=e^i$, so the consumer chooses $\mbx$ to maximize $V(\mbx,-\mathbf{p}^i\cdot \mbx)$.  If we are modeling price reference dependence, we could let $\phi (e,e^i)=ef(e/e^i)$, where $f(1)=1$.  The effective expenditure attributed to a bundle $\mbx$ is then $\mbp\cdot\mbx\, f((\mbp\cdot \mbx)/(\mbp^i\cdot\mbx))$.  If $f$ is increasing, then the consumer penalizes a bundle $\mbx$ when its actual cost $\mbp\cdot \mbx$ exceeds the benchmark cost $\mbp^i\cdot\mbx$ in the sense that the consumer attributes to it an expenditure $\mbp\cdot\mbx\, f((\mbp\cdot \mbx)/(\mbp^i\cdot\mbx))$ that is greater than the actual expenditure $\mbp\cdot\mbx$; similarly, a bundle that costs less than its benchmark is (psychologically) mapped to a cost that is even lower than the actual cost.

For a given $\phi$, we say that a collection of heterogeneous prices ${\cal P}= \{\mathbf{p}^{i,t}\}_{(i,t)\in N\times T}$ \textbf{AU-rationalizes} $\mathcal{E}=\big\{\big(\mathbf{e}^{i, t}\big)_{i\in N}, \overline{\mathbf{p}}^{t}\big\}_{t\in T}$ if there is an augmented utility function $V$ such that for each $(i, t)\in N\times T$,
$$V(\mbx^{i,t}, -\phi (\overline{\mbp}^t\cdot \mbx^{i,t},\mbp^{i,t}\cdot\mbx^{i,t}))\geq  V(\mbx, -\phi (\overline{\mbp}^t\cdot \mbx,\mbp^{i,t}\cdot\mbx ))\:\mbox{ for all $\mbx\in\mathbb{R}^K_+$,}$$
where $\mbx^{i,t}=\mbx(\mathbf{e}^{i,t},\overline{\mbp}^t)$.  The main result of this section establishes the equivalence between this notion of rationalizability and {\bf AU-rationalizability with heterogeneous preferences}, which means the following: there are augmented utility functions $V^1,V^2,\ldots ,V^N$ such that for each $(i, t)\in N\times T$,
$$V^i(\mbx^{i,t},-\overline{\mbp}^t\cdot \mbx^{i,t})\geq V^i(\mbx,-\overline{\mbp}^t\cdot\mbx)\:\mbox{ for all $\mbx\in\mathbb{R}^K_+$}.$$

We say that $\phi$ is \emph{regular} if
\begin{itemize}
\item[(a)]\, $\phi(e, \alpha\, e)$ is strictly increasing in $e$ for every $\alpha>0$;
\item[(b)]\, \emph{either} $\phi$ is increasing in its second argument and, for any compact set $I$ with $0\notin I$ such that $a, b, x\in I$, there is $m>0$ such that $\phi(a, x)>\phi(b, y)$ for any $y<m$ {\em or} $\phi$ is decreasing in its second argument and for any compact set $I$ with $0\notin I$ such that $a, b, y\in I$, there is $m>0$ such that $\phi(a, x)>\phi(b, y)$ for any $x<m$.
\end{itemize}
Condition (b) requires that if $x$ and $y$ are sufficiently far apart, then $\phi(a, x)-\phi(b, y)$ is positive.  Recall the two examples of $\phi$ we have discussed above. When $\phi(e, e')=e'$, $\phi$ is obviously regular.  If $\phi(e, e')=e\,f(e/e')$ with $f(1)=1$, then it clearly satisfies property (a).  If $f$ is increasing, continuous, and $\lim_{r\to \infty}f(r)=\infty$, then $\phi$ is decreasing in its second argument and satisfies property (b); if $f$ is decreasing, continuous, and $\lim_{r\to \infty}f(r)=0$, then $\phi$ is increasing in its second argument and satisfies property (b).   



The following result gives the precise sense in which preference and price heterogeneity have equivalent observable restrictions, in the context of the augmented utility model.   

\begin{theorem}\label{thm-six}
For any data set $\mathcal{E}=\big\{\big(\mathbf{e}^{i, t}\big)_{i\in N}, \overline{\mathbf{p}}^{t}\big\}_{t\in T}$ the following are equivalent.
\begin{itemize}
\item[{\em (i)}] $\cal E$ is AU-rationalizable with heterogeneous preferences.
\item[{\em (ii)}] Suppose $\phi$ is a regular behavioral expenditure function and $W$ is a linear homogeneous aggregator function satisfying  $\lim_{p^i\to \infty }W\big(p^i, (p^{j})_{j\neq i}\big)=\infty$ for any $(p^{j})_{j\neq i}\in\mathbb{R}^{N-1}_{++}$. Then there is a collection of prices ${\cal P}=\{\mathbf{p}^{i,t}\}_{(i,t)\in N\times T}$ that has a stable distribution (i.e., $\mathbf{p}^{i,t}=\lambda_i\,\overline{\mathbf{p}}^t$ for some $\lambda_i>0$), is consistent in expectation w.r.t. $W$, and AU-rationalizes $\cal E$ (for the given $\phi$).   
\end{itemize}
\end{theorem}


There is an analogous version of Theorem \ref{thm-six} in the setting of a cross-sectional data set. We say that a cross-sectional data set $\cal D$ is \textbf{random AU-rationalizable} (RAU-rationalizable) if there are sorting functions $\{\sigma^t\}_{t\in T}$ such that $\mathcal{E}(\{\sigma^t\}_{t\in T})=\big\{\big(\sigma^t(i)\big)_{i\in N}, \overline{\mathbf{p}}^{t}\big\}_{t\in T}$ is AU-rationalizable with heterogeneous preferences.\footnote{ \linespread{1.1} \selectfont There is a procedure to test for the RAU-rationalizability of a cross-sectional data set $\cal D$, which is similar to the test of RUM-rationalizability presented in \cite{kitamura2018nonparametric} (see \cite{deb2023revealed}).}  We can compare this notion of AU-rationalization with that under the \textbf{random price model} (RPM).  The cross-sectional data set $\cal D$ is  \textbf{AU-RPM-rationalizable} by a collection of heterogeneous prices ${\cal P}= \{\mathbf{p}^{i,t}\}_{(i,t)\in N\times T}$ if there are sorting functions $\{\sigma^t\}_{t\in T}$ such that $\mathcal{E}(\{\sigma^t\}_{t\in T})= \big\{\big(\sigma^t(i)\big)_{i\in N}, \overline{\mathbf{p}}^{t}\big\}_{t\in T}$ is AU-rationalized by $\cal P$. The following result is an immediate consequence of Theorem \ref{thm-six}.   


\begin{corollary}\footnote{\linespread{1.1} \selectfont It is possible to formulate a version of this theorem that allows for $\cal D$ to have non-discrete distributions.  Such a result would be analogous to the equivalence of RUM- and RPM-rationalizability when $\cal D$ has non-discrete distributions (see Theorem \ref{thm-nine} in the Appendix), and it will have a similar proof.  We will not present such a result since it does not convey a message not already conveyed by Corollary \ref{thm-seven}.}      \label{thm-seven} 
For any cross-sectional data $\mathcal{D}=\big\{E^t, \overline{\mathbf{p}}^{t}\big\}_{t\in T}$, the following are equivalent.\vspace{-0.05in}
\begin{itemize}
\item[{\em (i)}] $\mathcal{D}$ is RAU-rationalizable.\vspace{-0.05in}
\item[{\em (ii)}] Suppose $\phi$ is a regular behavioral expenditure function and $W$ is a linear homogeneous aggregator function satisfying  $\lim_{p^i\to \infty }W\big(p^i, (p^{j})_{j\neq i}\big)=\infty$ for any $(p^{j})_{j\neq i}\in\mathbb{R}^{N-1}_{++}$. Then $\cal D$ is AU-RPM rationalizable (for the given $\phi$) by some ${\cal P}=\{\mathbf{p}^{i,t}\}_{(i,t)\in N\times T}$, where $\cal P$ has a stable distribution and is consistent in expectation w.r.t. $W$. 
\end{itemize}
\end{corollary}

To prove Theorem \ref{thm-six}, we use Theorem 2 of \cite{deb2023revealed}, which characterizes a more general version of the AUM that allows for non-linear pricing. Consider a collection of functions (or price systems) $\{f^t\}_{t\in T}$ on $\mathbb{R}^K_{+}$, where $f^t(\mathbf{x})$ is interpreted as the cost of purchasing $\mathbf{x}$ at observation $t$. For example, if prices are linear and $\overline{\mathbf{p}}^t$ are the prevailing prices, then $f^t(\mathbf{x})=\overline{\mathbf{p}}^t\cdot \mathbf{x}$. We say a collection of consumption bundles and price systems $\{(\mathbf{x}^t, f^t)\}_{t\in T}$ is \emph{AU-rationalizable} if there is an augmented utility function $V$ such that for each $t\in T$,
$$V(\mbx^{t},-f^t(\mbx^{t}))\geq V(\mbx,-f^t(\mbx))\:\mbox{ for all $\mbx\in\mathbb{R}^K_+$}.$$
Abusing notation somewhat, we say that a collection of consumption bundles and price vectors $\{(\mathbf{x}^t, \overline{\mathbf{p}}^t)\}_{t\in T}$ is AU-rationalizable if it is AU-rationalizable with $f^t$ given by $f^t(\mathbf{x})=\overline{\mathbf{p}}^t\cdot\mathbf{x}$.   

Theorem 2 of \cite{deb2023revealed} shows that AU-rationalization is characterized by a type of no-cycling condition, which we call \emph{generalized axiom of price preference} (GAPP). We say $f^t$ is \textbf{directly revealed (strictly) preferred to} $f^s$, denoted by $f^t\succsim_{P} f^s$, if $f^t(\mathbf{x}^s)\le (<) f^s(\mathbf{x}^s)$. We say $f^t$ is \textbf{revealed preferred to} $f^s$, denoted by $f^t\succsim^*_{P} f^s$, if there is a sequence $\{f^{t_l}\}^L_{l=1}$ such that $f^{t_1}=f^t$, $f^{t_L}=f^s$, and $f^{t_l}\succsim_{P} f^{t_{l+1}}$ for each $l<L$. We say that $\{(\mathbf{x}^t, f^t)\}_{t\in T}$ satisfies GAPP if, for any $s, t\in T$, $f^t\succsim^*_{P} f^s$ implies $f^s\not\succ_{P} f^t$; GAPP is a necessary and sufficient condition for $\{(\mathbf{x}^t,f^t)\}_{t\in T}$ to be AU-rationalizable.  


\medskip
\noindent\textbf{Proof sketch of Theorem \ref{thm-six}.} The detailed proof can be found in the Appendix, but the following quick sketch may be helpful.  Fix $\lambda_1, \ldots, \lambda_N$ and let $f^{i,t}(\mathbf{x})\equiv\phi(\overline{\mathbf{p}}^t \cdot \mathbf{x}, \lambda_i\,\overline{\mathbf{p}}^t\cdot\mathbf{x})$. Then by Theorem 2 of \cite{deb2023revealed}, $\mathcal{P}$ AU-rationalizes $\mathcal{E}$ iff $\{(\mathbf{x}^{i, t}, f^{i, t})\}_{(i, t)\in N\times T}$ satisfies GAPP. On the other hand, $\cal E$ is AU-rationalizable with heterogeneous preferences iff $\{(\mathbf{x}^{i, t}, \overline{\mathbf{p}}^{t})\}_{t\in T}$ satisfies GAPP for each $i$. It is now easy to see that $(ii)$ implies $(i)$.  Indeed, if $\{(\mathbf{x}^{i, t}, f^{i, t})\}_{(i, t)\in N\times T}$ satisfies GAPP, then $\{(\mathbf{x}^{i, t}, f^{i, t})\}_{t\in T}$ satisfies GAPP for each $i$.
Furthermore, $\phi$ is regular and thus satisfies property (a), which implies that $f^t(\mathbf{x}^s)\le (<)\, f^s(\mathbf{x}^s)$ if and only if $\overline{\mathbf{p}}^t\cdot \mathbf{x}^s\leq (<)\, \overline{\mathbf{p}}^s\cdot \mathbf{x}^s$.  We conclude that $\{(\mathbf{x}^{i, t}, f^{i, t})\}_{t\in T}$ satisfies GAPP if and only if $\{(\mathbf{x}^{i, t}, \overline{\mathbf{p}}^{t})\}_{t\in T}$ satisfies GAPP.

To prove that (i) implies (ii), we construct $\lambda_1, \ldots, \lambda_N$ such that $\lambda_i/\lambda_{i+1}$ is sufficiently large (or small) so that $f^{j, s}$ cannot be revealed preferred to $f^{i, t}$ when $i<j$; in other words, there will be no revealed preference cycle involving observations from different agents. We show how this can be done in the Appendix, relying on the regularity of $\phi$ (and property (b) in particular).  Therefore, checking whether $\{(\mathbf{x}^{i, t}, f^{i, t})\}_{(i, t)\in N\times T}$ satisfies GAPP is equivalent to checking whether $\{(\mathbf{x}^{i, t}, f^{i, t})\}_{t\in T}$ satisfies GAPP for each $i$.  By property (a) of the regularity of $\phi$, $\{(\mathbf{x}^{i, t}, f^{i, t})\}_{t\in T}$ satisfies GAPP if and only if  $\{(\mathbf{x}^{i, t}, \overline{\mathbf{p}}^t)\}_{t\in T}$ satisfies GAPP. Lastly, the latter property holds since (i) holds. \hfill \qed

\section*{Appendix}

\noindent {\bf Proof of Proposition \ref{One}.}\,  By Afriat's theorem, we shall construct heterogeneous prices $\mathcal{P}=\{\mathbf{p}^{i, t}\}_{(i, t)\in N\times T}$, which are consistent in expectation and satisfy $p^{i,t}_k=\overline{p}_k^t$ for all $k>2$, such that a data set $\mathcal{O}^*=\{(\mathbf{x}(\mathbf{e}^{i, t}, \mathbf{p}^{i, t}), \mathbf{p}^{i, t})\}_{(i, t)\in N\times T}$ satisfies GARP. For any $(i, t)\in N\times T$ and $k\in K$ with $k\ge 3$, let $p^{i, t}_k\equiv \overline{p}^t_k$. Hence, $W^t_k\big(p^{1, t}_k, \ldots, p^{N, t}_k\big)=W^t_k\big(\overline{p}^{t}_k, \ldots, \overline{p}^{t}_k\big)=\overline{p}^{t}_k$ when $k\ge 3$.

Let $>^*$ be a lexicographic order on $N\times T$ such that $(j, s)>^* (i, t)$ if $j>i$ or $i=j$ and $s>t$. In other words, $(i, t+1)>^*(i, t)$ and $(i+1, 1)>^*(i, T)$. We now construct $\{p^{i, t}_1, p^{i, t}_2\}_{(i, t)\in N\times T}$ such that the bundle $\mathbf{x}(\mathbf{e}^{j, s}, \mathbf{p}^{j, s})$ for $(j, s)$ is not affordable for $(i, t)$ whenever $(j, s)>^*(i, t)$. That is, $\mathbf{x}(\mathbf{e}^{i, t}, \mathbf{p}^{i, t})\not\succsim_D \mathbf{x}(\mathbf{e}^{j, s}, \mathbf{p}^{j, s})$ (i.e., $\mathbf{p}^{i, t} \cdot \mathbf{x}(\mathbf{e}^{j, s}, \mathbf{p}^{j, s})>\mathbf{p}^{i, t} \cdot \mathbf{x}(\mathbf{e}^{i, t}, \mathbf{p}^{i, t})=m^{i, t}$) when $(j, s)>^* (i, t)$.

We first construct $p^{i, t}_1$ with $i<N$ as follows. Take any $p_1>0$ and let $p^{1, 1}_1\equiv p_1$. Moreover, take any small enough $\epsilon>0$ and we can construct the rest of  $p^{i, t}_1$ such that \[\epsilon\, p^{i, t}_1\ge p^{j, s}_1\text{ whenever }(j, s)>^*(i, t).\]
In words, the price of good 1 is lower in later periods and is also lower for agents with higher indexes; i.e., $p^{i, t}_1$ is decreasing exponentially in $i$ as well as $t$. In particular, set $p^{i, t}_1=p_1\,\epsilon^{(i-1)T+t-1}$. Therefore, $\mathbf{x}^{j, s}(\mathbf{e}^{j, s}, \mathbf{p}^{j, s})$ is not affordable for $(i, t)$ when $(j, s)>^*(i, t)$ and $i, j<N$. More formally, when $i, j<N$, when $\epsilon$ is small enough we have\[\mathbf{p}^{i, t} \cdot \mathbf{x}(\mathbf{e}^{j, s}, \mathbf{p}^{j, s})=\sum_{k\in K}p^{i, t}_k\,\frac{e^{j, s}_k}{p^{j, s}_k}>\frac{p^{i, t}_{1}}{p^{j, s}_{1}}\,e^{j, s}_{1}\ge\frac{e^{j, s}_{1}}{\epsilon}>m^{i,t}\text{ since }\frac{p^{i, t}_{1}}{p^{j, s}_{1}}\ge \frac{1}{\epsilon}.\]

Second, let $p^{N, t}_1$ be the solution to $W^t_1\big(p^{1, t}_1, \ldots, p^{N-1, t}_1, p^{N, t}_1\big)=\overline{p}^{t}_1$. Third, take any $p_2>0$ and let $p^{N, t}_2\equiv \epsilon^{t} p_2$. Similar to the previous argument, the bundle $\mathbf{x}(\mathbf{e}^{N, t+1}, \mathbf{p}^{N, t+1})$ for $(N, t+1)$ is not affordable for $(N, t)$ (i.e., $\mathbf{x}(\mathbf{e}^{N, t}, \mathbf{p}^{N, t})\not\succsim_D \mathbf{x}(\mathbf{e}^{N, t+1}, \mathbf{p}^{N, t+1})$) since the price of good 2 for agent $N$ is exponentially decreasing in $t$. Fourth, for any $(i, t)$ with $i<N$, let $p^{i, t}_2\equiv \tilde{p}^t_2$ where $\tilde{p}^t_2$ solves $W^t_2\big(\tilde{p}^{t}_2, \ldots, \tilde{p}^{t}_2, \epsilon^{t}\,p_2\big)=\overline{p}^{t}_2$.

\medskip
\noindent\textbf{Case 1.} Suppose for any $(p^{j})_{j\neq i}\in\mathbb{R}^{N-1}_{++}$ and $(t, k)\in T\times K$, $\lim_{p^i\to+\infty }W^t_k\big(p^i, (p^{j})_{j\neq i}\big)=+\infty$.\smallskip

In this case, the equation $W^t_1\big(p^{1, t}_1, \ldots, p^{n-1, t}_1, p^{N, t}_1\big)=\overline{p}^{t}_1$ has a solution when $p_1$ is small enough so that $\overline{p}^t_1\ge p^{i, t}_1=p_1\,\epsilon^{(i-1)T+t-1}$ for each $i<N$. Moreover, $W^t_2\big(\tilde{p}^{t}_2, \ldots, \tilde{p}^{t}_2, \epsilon^{t}\,p_2\big)=\overline{p}^{t}_2$ has a solution when $p_2$ is small enough.

To prove GARP, it is sufficient to show that $\mathbf{x}(\mathbf{e}^{i, t}, \mathbf{p}^{i, t})\not\succsim_D \mathbf{x}(\mathbf{e}^{j, s}, \mathbf{p}^{j, s})$ whenever $(j, s)>^*(i, t)$. By the previous arguments, it is enough to consider the case in which $j=N$ and $i<N$. Since $p^{i, t}_2=\tilde{p}^{t}_{2}\ge \overline{p}^t_2$, we have $\mathbf{x}(\mathbf{e}^{i, t}, \mathbf{p}^{i, t})\not\succsim_D \mathbf{x}(\mathbf{e}^{N, s}, \mathbf{p}^{N, s})$; i.e.,
\[\mathbf{p}^{i, t} \cdot \mathbf{x}(\mathbf{e}^{N, s}, \mathbf{p}^{N, s})=\sum_{k\in K}p^{i, t}_k\,\frac{e^{N, s}_k}{p^{N, s}_k}>\frac{p^{i, t}_{2}}{p^{N, s}_{2}}\,e^{N, s}_{2}
\ge\frac{\overline{p}^{t}_{2}}{\epsilon^s\, p_2}\,e^{N, s}_{2}>m^{i, t}\text{ when $\epsilon$ and $p_2$ are small enough}.\]

\noindent\textbf{Case 2.} Suppose for any $(p^{j})_{j\neq i}\in\mathbb{R}^{N-1}_{++}$ and $(t, k)\in T\times K$, $\lim_{p^i\to+0}W^t_k\big(p^i, (p^{j})_{j\neq i}\big)=0$.

In this case, the equation $W^t_1\big(p^{1, t}_1, \ldots, p^{n-1, t}_1, p^{N, t}_1\big)=\overline{p}^{t}_1$ has a solution when $p_1$ is large enough so that $\frac{\overline{p}^t_1}{\epsilon}\le p^{i, t}_1=p_1\,\epsilon^{(i-1)T+t-1}$ for each $i<N$. Moreover, $W^t_2\big(\tilde{p}^{t}_2, \ldots, \tilde{p}^{t}_2, \epsilon^{t}\,p_2\big)=\overline{p}^{t}_2$ has a solution when $p_2$ is large enough.

To prove GARP, it is sufficient to show that $\mathbf{x}(\mathbf{e}^{i, t}, \mathbf{p}^{i, t})\not\succsim_D \mathbf{x}(\mathbf{e}^{j, s}, \mathbf{p}^{j, s})$ whenever $(j, s)>^*(i, t)$. By the previous arguments, it is enough to consider the case in which $j=N$ and $i<N$. Since $p^{i, t}_{1}\ge \frac{\overline{p}^{t}_{1}}{\epsilon}$ and $p^{N, s}_1\le \overline{p}^s_1$, we have $\mathbf{x}(\mathbf{e}^{i, t}, \mathbf{p}^{i, t})\not\succsim_D \mathbf{x}(\mathbf{e}^{N, s}, \mathbf{p}^{N, s})$; i.e.,
\[\mathbf{p}^{i, t} \cdot \mathbf{x}(\mathbf{e}^{N, s}, \mathbf{p}^{N, s})=\sum_{k\in K}p^{i, t}_k\,\frac{e^{N, s}_k}{p^{N, s}_k}>\frac{p^{i, t}_{1}}{p^{N, s}_{1}}\,e^{N, s}_{1}\ge\frac{\overline{p}^{t}_{1}}{\epsilon\,\overline{p}^s_1}\,e^{N, s}_{1}>m^{i, t}\]
when $\epsilon$ is small enough.\hfill \qed\medskip

Below, we provide an example of a data set that cannot be rationalized by price heterogeneity in one good.  This shows that, in Proposition \ref{One}, it is essential that there be at least two goods with price heterogeneity.  In fact, the example establishes a stronger conclusion: the data set it considers cannot be rationalized by price heterogeneity in a single good alone, even if we allow the two agents to have distinct preferences.\medskip

\noindent {\bf Example.}\, We assume that there are two agents, four goods, and four observations.  The two agents have identical expenditure allocations, and the observations are as follows.
\begin{center}
\begin{tabular}{cccc}
$\mathbf{e}^{i, 1}=(10, 1, 1, 1)$ & $\mathbf{e}^{i, 2}=(1, 10, 1, 1)$ & $\mathbf{e}^{i, 3}=(1, 1, 10, 1)$ & $\mathbf{e}^{i, 4}=(1, 1, 1, 10)$  \\
$\overline{\mathbf{p}}^{1}=(2, 1, 1, 1)$ & $\overline{\mathbf{p}}^{2}=(1, 2, 1, 1)$ & $\overline{\mathbf{p}}^{3}=(1, 1, 2, 1)$ & $\overline{\mathbf{p}}^{4}=(1, 1, 1, 2)$
\end{tabular}
\end{center}
Proposition \ref{One} guarantees that this data can be rationalized by heterogeneous prices with consistent expectations, so long as there is price heterogeneity in at least two goods.  However, assuming that prices are aggregated by taking the arithmetic average, we claim that this data  {\em cannot} be rationalized with price heterogeneity in good 1 alone, even if we allow the two agents to have different preferences.  Of course, given the symmetry, it is clear that this data set cannot be rationalized by price heterogeneity in any other good alone.

Let $x^{i, t}_k= e^{i, t}_k/p^{i, t}_k$. Note that for any $i\in N$ and distinct $t, t'\in \{2, 3, 4\}$, $\mathbf{x}^{i, t}$ is directly revealed strictly preferred to $\mathbf{x}^{i, t'}$ if and only if
$$p^{i,t}_1\times\frac{1}{p^{i,t'}_1}+1\times 1+1\times 5+2\times1<13$$
or $p^{i,t}_1/p^{i,t'}_1<5$. Therefore, $\mathbf{x}^{i, t}$ is directly revealed strictly preferred to $\mathbf{x}^{i, t'}$ and vice versa if and only if
\begin{equation}\label{violateG}
\frac{1}{5}<\frac{p^{i,t'}_1}{p^{i,t}_1}<5.
\end{equation}
Since $\frac{p^{1,t}_1+p^{2, t}_1}{2}=1$ for any $t=2,3,4$, there is an agent $i$ and distinct $t, t'\in \{2, 3, 4\}$ such that $2\geq p^{i, t}, p^{i, t'}\geq 1$.  Therefore,
$$\frac{1}{2}<\frac{p^{i,t}_1}{p^{i,t'}_1}<2$$
which means (given (\ref{violateG})) that agent $i$ has violated GARP.  \hfill \qed
\medskip

\noindent \textbf{Proof of Proposition \ref{prop-indistinct}.}\, Take any stable price distribution ${\cal P}=\{\mathbf{p}^{i,t}\}_{(i,t)\in N\times T}$ where $p^{i, t}_k=\lambda_{i,k}\,\overline{p}^t_k$ for some $\{\lambda_{i,k}\}_{(i, k)\in N\times K}$. By Afriat's Theorem, $\mathcal{E}=\{(\mathbf{e}^{i, t})_{i\in N}, \overline{\mathbf{p}}^{t}\}_{t\in T}$ is rationalizable with heterogeneous preferences if and only if $\overline{\mathcal{O}}^i=\{(\mathbf{x}(\mathbf{e}^{i,t}, \overline{\mathbf{p}}^{t}), \overline{\mathbf{p}}^{t})\}_{t\in T}$ satisfies GARP for each $i\in N$. Similarly, $\{(\mathbf{e}^{i, t}, \mathbf{p}^{i, t})\}_{(i, t)\in N\times T}$ is rationalizable with heterogeneous preferences if and only if $\mathcal{O}^i=\{(\mathbf{x}(\mathbf{e}^{i,t}, \mathbf{p}^{i,t}), \mathbf{p}^{i,t})\}_{t\in T}$ satisfies GARP for each $i\in N$. To establish our result, we need only show that for any $i\in N$, $\mathcal{O}^i$ satisfies GARP if and only if $\overline{\mathcal{O}}^i$ satisfies GARP. Indeed, note that the revealed preference relations on $\mathcal{O}^i$ and $\overline{\mathcal{O}}^i$ are the `same' in the following sense: for any $s, t\in T$, $\mathbf{x}(\mathbf{e}^{i, t}, \mathbf{p}^{i, t})$ is directly revealed preferred to $\mathbf{x}(\mathbf{e}^{i, s}, \mathbf{p}^{i, s})$ if and only if $\mathbf{x}(\mathbf{e}^{i, t}, \overline{\mathbf{p}}^{t})$ is directly revealed preferred to $\mathbf{x}(\mathbf{e}^{i, s}, \overline{\mathbf{p}}^{s})$. This is simply because, by the definition of revealed preference relations, $\mathbf{x}(\mathbf{e}^{i, t}, \mathbf{p}^{i, t})\succsim_{D}\mathbf{x}(\mathbf{e}^{i, s}, \mathbf{p}^{i, s})$ is equivalent to
\[\mathbf{p}^{i, t}\cdot \mathbf{x}(\mathbf{e}^{i, t}, \mathbf{p}^{i, t})=m^{i, t}\ge \mathbf{p}^{i, t}\cdot \mathbf{x}(\mathbf{e}^{i, s}, \mathbf{p}^{i, s})=\sum_{k\in K}p^{i, s}_k\frac{e^{i, t}_k}{p^{i, t}_k}=\sum_{k\in K}\lambda_{i,k}\,\overline{p}^{s}_k\frac{e^{i, t}_k}{\lambda_{i,k}\,\overline{p}^{t}_k}=\sum_{k\in K}\overline{p}^{s}_k\frac{e^{i, t}_k}{\overline{p}^{t}_k}\]
while $\mathbf{x}(\mathbf{e}^{i, t}, \overline{\mathbf{p}}^{t})\succsim_{D}\mathbf{x}(\mathbf{e}^{i, s}, \overline{\mathbf{p}}^{s})$ is also equivalent to
\[\overline{\mathbf{p}}^{t}\cdot \mathbf{x}(\mathbf{e}^{i, t}, \overline{\mathbf{p}}^{t})=m^{i, t}\ge \overline{\mathbf{p}}^{t}\cdot \mathbf{x}(\mathbf{e}^{i, t}, \overline{\mathbf{p}}^{s})=\mathbf{x}(\mathbf{e}^{i, t}, \mathbf{p}^{i, t})=\sum_{k\in K}\overline{p}^{s}_k\frac{e^{i, t}_k}{\overline{p}^{t}_k}.\]
Hence, for each $i\in N$, $\mathcal{O}^i$ satisfies GARP if and only if $\overline{\mathcal{O}}^i$ satisfies GARP.
\hfill \qed

\noindent {\bf Proof of Theorem \ref{thm-four}.}\,  Clearly, (ii) implies (i).  To show that (iii) implies (ii), take any $R\subseteq K$ with $\text{e}^{i, t}_R>0$ for each $(i, t)\in N\times T$. Suppose there are real numbers $\lambda_1, \ldots, \lambda_N>0$ such that stable price distribution ${\cal P}=\{\mathbf{p}^{i,t}\}_{(i,t)\in N\times T}$ with $\mathbf{p}^{i, t}=(\lambda_i\, \overline{\mathbf{p}}^t_R, \overline{\mathbf{p}}^t_{-R})$ rationalizes $\mathcal{E}$.\footnote{\linespread{1.1} \selectfont We will not use the fact that prices are consistent in expectation.} Hence, there is a utility function $U:\mathbb{R}^K_{+}\to\mathbb{R}$ such that for each $(i, t)\in N\times T$,
\begin{equation}\label{seventeen}
U(\mathbf{x}(\mathbf{e}^{i,t}, \mathbf{p}^{i,t})) \ge U(\mathbf{x}) \text{ for any }\mathbf{x}\in \mathscr{B}\left(\mathbf{p}^{i,t}, m^{i, t}\right).\end{equation}
Since $\mathbf{x}_R(\mathbf{e}^{i,t}, \mathbf{p}^{i,t})=\mathbf{x}_R(\mathbf{e}^{i,t}, \overline{\mathbf{p}}^t)/\lambda_i$ and $\mathbf{x}_{-R}(\mathbf{e}^{i,t}, \mathbf{p}^{i,t})=\mathbf{x}_{-R}(\mathbf{e}^{i,t}, \overline{\mathbf{p}}^t)$, we have
$$U(\mathbf{x}(\mathbf{e}^{i,t}, \mathbf{p}^{i,t}))=U\left(\frac{\mathbf{x}_R(\mathbf{e}^{i,t}, \overline{\mathbf{p}}^t)}{\lambda_i}, \mathbf{x}_{-R}(\mathbf{e}^{i,t}, \overline{\mathbf{p}}^t)\right)=U^{\frac{1}{\lambda_i}}_R(\mathbf{x}(\mathbf{e}^{i,t}, \overline{\mathbf{p}}^t)).$$
Let $\tilde{\mathbf{x}}=(\lambda_i \mathbf{x}_R, \mathbf{x}_{-R})$. Then $U(\mathbf{x})=U^{\frac{1}{\lambda_i}}_R(\lambda_i \mathbf{x}_R, \mathbf{x}_{-R})=U^{\frac{1}{\lambda_i}}_R(\tilde{\mathbf{x}})$. Moreover,
\[\mathbf{x}\in \mathscr{B}\left(\mathbf{p}^{i,t}, m^{i, t}\right)\text{ if and only if }\tilde{\mathbf{x}}\in\mathscr{B}\left(\overline{\mathbf{p}}^{t}, m^{i, t}\right)\]
since $\mathbf{p}^{i,t}\cdot\mathbf{x}=\lambda_i\,\overline{\mathbf{p}}^t_R\cdot\mathbf{x}_R+\overline{\mathbf{p}}^t_{-R}\cdot\mathbf{x}_{-R}
=\overline{\mathbf{p}}^t\cdot \tilde{\mathbf{x}}$. Finally, the above three observations show that (\ref{seventeen}) is equivalent to
\[U^{\frac{1}{\lambda_i}}_R(\mathbf{x}(\mathbf{e}^{i,t}, \overline{\mathbf{p}}^t))\ge U^{\frac{1}{\lambda_i}}_R(\tilde{\mathbf{x}})\text{ for any }\tilde{\mathbf{x}}\in\mathscr{B}\left(\overline{\mathbf{p}}^{t}, m^{i, t}\right).\]
Therefore, $\cal E$ is rationalizable with heterogeneous preferences drawn from an $R$-scale family.

It remains to show that (i) implies (iii).  Suppose $\mathcal{E}$ is rationalizable with heterogeneous preferences. We shall find $\lambda_1, \ldots, \lambda_N>0$ such that stable price distribution ${\cal P}=\{\mathbf{p}^{i,t}\}_{(i,t)\in N\times T}$ with $\mathbf{p}^{i, t}=(\lambda_i\, \overline{\mathbf{p}}^t_R, \overline{\mathbf{p}}^t_{-R})$ is consistent in expectation and that rationalizes $\mathcal{E}$.

Let $\lambda_i=\beta\,\epsilon^i$ where $\epsilon>0$ and $\beta=1/W((\epsilon^i)_{i\in N})$. By homogeneous of degree 1 of $W$, we have $W((\lambda_i\,\overline{p}^{t}_k)_{i\in N})=W((\beta\,\epsilon^i\,\overline{p}^{t}_k)_{i\in N})=\overline{p}^{t}_k\,\beta\,W((\epsilon^i)_{i\in N})=\overline{p}^{t}_k$ for any $k\in R$ and $t\in T$. By the property of $W$, we also have $W((p^{i, t}_k)_{i\in N})=W((\overline{p}^{t}_k)_{i\in N})=\overline{p}^{t}_k$ for any $k\not\in R$ and $t\in T$. Therefore, ${\cal P}=\{\mathbf{p}^{i,t}\}_{(i,t)\in N\times T}$ is consistent in expectation.

We now shall show that ${\cal P}=\{\mathbf{p}^{i,t}\}_{(i,t)\in N\times T}$ rationalizes $\mathcal{E}$ when $\epsilon$ is small enough. By Afriat's theorem, it is enough to show that a data set $\mathcal{O}^*=\left\{\left(\mathbf{x}(\mathbf{e}^{i,t}, \mathbf{p}^{i,t}),\mathbf{p}^{i,t}\right)\right\}_{(i,t)\in N\times T}$ obeys GARP when $\epsilon$ is small enough.  We write $\mathcal{O}^*=\bigcup_{i\in N }\mathcal{O}^i$, where $\mathcal{O}^i=\{(\mathbf{x}(\mathbf{e}^{i, t}, \mathbf{p}^{i, t}), \mathbf{p}^{i, t})\}_{t\in T}$.

Let $\succsim_{\overline{D}}$ be the revealed preference relation on $\mathcal{O}^*$. We claim that when $\epsilon$ is sufficiently small, for any $i, j\in N$ with $i>j$ and $s, t\in T$, $\mathbf{x}(\mathbf{e}^{j, s}, \mathbf{p}^{j, s})\not\succsim_{\overline{D}} \mathbf{x}(\mathbf{e}^{i, t}, \mathbf{p}^{i, t})$. Since
\[\mathbf{p}^{j, s}\cdot \mathbf{x}(\mathbf{e}^{i, t}, \mathbf{p}^{i, t})=\frac{\lambda_j}{\lambda_i}\sum_{k\in R}\overline{p}^{s}_k\,\frac{e^{i, t}_k}{\overline{p}^{t}_k}+\sum_{k\not\in R}\overline{p}^{s}_k\frac{e^{i, t}_k}{\overline{p}^{t}_k}\ge \frac{1}{\epsilon^{i-j}} \,\sum_{k\in R}\overline{p}^{s}_k\,\frac{e^{i, t}_k}{\overline{p}^{t}_k}\]
and $\mathbf{e}^{i, t}_R>0$, $\mathbf{p}^{j, s}\cdot \mathbf{x}(\mathbf{e}^{i, t}, \mathbf{p}^{i, t})$ will exceed $\mathbf{p}^{j, s}\cdot \mathbf{x}(\mathbf{e}^{j, s}, \mathbf{p}^{j, s})=\sum_{k\in K}e_k^{j,s}$ when $\epsilon$ is small enough and $i>j$.  Therefore, there is no revealed preference cycle involving observations from ${\cal O}^i$ and ${\cal O}^j$ when $i\neq j$. Lastly, since $\mathcal{E}$ is rationalizable with heterogeneous preferences, Proposition \ref{prop-indistinct} guarantees that ${\cal O}^i$ satisfies GARP for each $i\in N$.   Therefore, $\mathcal{O}^*$ satisfies GARP.\hfill \qed\medskip

\noindent {\bf Proof of Theorem \ref{thm-six}.}\, We have already shown that $(ii)$ implies $(i)$ in the proof sketch provided in Section \ref{sec5-result}.  It remains for us to show that $(i)$ implies $(ii)$.  First, observe that, for any $\lambda_i>0$, $\{(\mathbf{x}^{i,t},f^{i,t})\}_{t\in T}$ satisfies GAPP. Indeed, if $(i)$ holds, then $\{(\mathbf{x}^{i,t},\overline{\mathbf{p}}^{t})\}_{t\in T}$ satisfies GAPP and then the regularity of $\phi$ (specifically,  property (a)) guarantees that  
$\{(\mathbf{x}^{i,t},f^{i,t})\}_{t\in T}$ also satisfies GAPP (see the argument in the proof sketch). Furthermore, we claim that we can construct $\lambda_1, \ldots, \lambda_N$ such that there are no revealed preference cycles involving $f^{i,t}$ and $f^{j,s}$ with $i\neq j$; we establish this claim below using property (b) in the definition of the regularity of $\phi$. We can thus conclude that $\{(\mathbf{x}^{i, t}, f^{i, t})\}_{(i, t)\in N\times T}$ satisfies GAPP and, by Theorem 2 of \cite{deb2023revealed}, it is AU-rationalizable, as required by $(ii)$. We break down our proof into two cases, corresponding to the two conditions required for the regularity of $\phi$ under property (b).  

\smallskip
\noindent\textbf{Case 1.} $\phi$ is increasing in its second argument and, for any compact set $I$ with $0\notin I$ such that $a, b, x\in I$, there is $m>0$ such that $\phi(a, x)>\phi(b, y)$ for any $y<m$.\smallskip

It is straightforward to show that this combination of assumptions on $\phi$ guarantees that for any $a$, $b\in I$ such that $I$ is a compact set with $0\notin I$, and $x\in S$ where $S$ is uniformly bounded away from zero, there is $m>0$ such that $\phi(a, x)>\phi(b, y)$ for any $y<m$.   It is this property on $\phi$ specifically that we need in our proof.     

We now explain how we can construct positive numbers $\lambda_{N-1},\ldots,\lambda_2,\lambda_1$ such that, for any $j>i$ and $s, t\in T$, $f^{j, s}\not\succsim_{P} f^{i, t}$ for any $\lambda_N\geq 1$. First, note that $\{\overline{\mathbf{p}}^s\cdot\mathbf{x}^{N-1,t}\}_{s,t\in T}$ is a compact set not containing zero, and the set $\{\lambda_N\overline{\mathbf{p}}^s\cdot\mathbf{x}^{N-1,t}\}_{s, t\in T,\, \lambda_N\geq 1}$ is uniformly bounded away from zero. Therefore, there is $0<\lambda_{N-1}<1$ such that 
\[f^{N, s}(\mathbf{x}^{N-1, t})=\phi(\overline{\mathbf{p}}^s\cdot \mathbf{x}^{N-1, t}, \lambda_N\,\overline{\mathbf{p}}^s\cdot \mathbf{x}^{N-1, t})>f^{N-1, t}(\mathbf{x}^{N-1, t})=\phi(\overline{\mathbf{p}}^t\cdot \mathbf{x}^{N-1, t}, \lambda_{N-1}\,\overline{\mathbf{p}}^t\cdot \mathbf{x}^{N-1, t})\]
when $\lambda_{N-1}$ is small enough.  Note that $\lambda_{N-1}$ can be chosen independently of the value of $\lambda_N$. Assuming that we have chosen $\lambda_{N-1},\ldots,\lambda_{i+1}$, we now explain how $\lambda_i$ can be chosen.  Observe that $\{\overline{\mathbf{p}}^s\cdot\mathbf{x}^{i,t}\}_{s,t\in T}$ is a compact set not containing zero and the set  
$\{\lambda_N\,\overline{\mathbf{p}}^s\cdot\mathbf{x}^{i,t}\}_{s, t\in T,\, \lambda_N\geq 1}\cup \{\lambda_j\,\overline{\mathbf{p}}^s\cdot\mathbf{x}^{i,t}\}_{s, t\in T,\, N>j>i}$ is uniformly bounded away from zero. Therefore, there is $\lambda_i\in (0,1)$ such that for all $j>i$, 
\[f^{j, s}(\mathbf{x}^{i, t})=\phi(\overline{\mathbf{p}}^s\cdot \mathbf{x}^{i, t}, \lambda_j\,\overline{\mathbf{p}}^s\cdot \mathbf{x}^{i, t})>f^{i, t}(\mathbf{x}^{i, t})=\phi(\overline{\mathbf{p}}^t\cdot \mathbf{x}^{i, t}, \lambda_i\,\overline{\mathbf{p}}^t\cdot \mathbf{x}^{i, t})\]
when $\lambda_i$ is sufficiently small.  Note that $\lambda_i$ can be chosen independently of $\lambda_N$. 

We have shown that there is $\lambda_{N-1}, \ldots, \lambda_2, \lambda_1$ such that for any $\lambda_N\geq 1$ and $i, j\in N$ with $i<j$ and $s, t\in T$, $f^{j, s}\not\succsim_{P} f^{i, t}$. Thus there can be no revealed preference cycles involving $f^{i,t}$ and $f^{j,s}$ with $i\neq j$. We also know that $W(\lambda_1,\lambda_2,\ldots,1)\leq 1$, so for $\lambda_N$ sufficiently large,  $W(\lambda_1,\lambda_2,\ldots,\lambda_N)= 1$, as required.  

\noindent\textbf{Case 2.} $\phi$ is decreasing in its second argument, and for any compact set $I$ with $0\notin I$ such that $a, b, y\in I$, there is $m>0$ such that $\phi(a, x)>\phi(b, y)$ for any $x<m$.\smallskip

The proof is analogous to the proof for Case 1. Note that the combination of assumptions on $\phi$ in Case 2 guarantees that for any $a$, $b\in I$ such that $I$ is a compact set with $0\notin I$ and $y\in S$ where $S$ is uniformly bounded away from zero, there is $m>0$ such that $\phi(a, x)>\phi(b, y)$ for any $x<m$.   It is this property on $\phi$ specifically that we need in our proof.   

We now explain how we can construct positive numbers $\lambda_{N-1},\ldots,\lambda_2,\lambda_1$ such that, for any $j>i$ and $s, t\in T$, $f^{i, s}\not\succsim_{P} f^{j, t}$ for any $\lambda_N\geq 1$. First, note that $\{\overline{\mathbf{p}}^s\cdot\mathbf{x}^{N,t}\}_{s,t\in T}$ is a compact set bounded away from zero and the set $\{\lambda_N\overline{\mathbf{p}}^t\cdot\mathbf{x}^{N,t}\}_{t\in T,\, \lambda_N\geq 1}$ is uniformly bounded away from zero. Therefore, there is $0<\lambda_{N-1}<1$ such that 
\[f^{N, t}(\mathbf{x}^{N, t})=\phi(\overline{\mathbf{p}}^t\cdot \mathbf{x}^{N, t}, \lambda_N\,\overline{\mathbf{p}}^t\cdot \mathbf{x}^{N, t}) < f^{N-1, s}(\mathbf{x}^{N, t})=\phi(\overline{\mathbf{p}}^s\cdot \mathbf{x}^{N, t}, \lambda_{N-1}\,\overline{\mathbf{p}}^s\cdot \mathbf{x}^{N, t})\]
when $\lambda_{N-1}$ is small enough.  Note that $\lambda_{N-1}$ can be chosen independently of the value of $\lambda_N$.  So this guarantees that 
$f^{N-1, s}\not\succsim_{P} f^{N, t}$ for any $\lambda_N\geq 1$. Assuming that we have chosen $\lambda_{N-1},\ldots,\lambda_{i+1}$, we now explain how $\lambda_i$ can be chosen. Observe that $\{\overline{\mathbf{p}}^s\cdot\mathbf{x}^{j,t}\}_{s,t\in T,\,j>i}$ is a compact set not containing zero and the set  
$\{\lambda_N\,\overline{\mathbf{p}}^t\cdot\mathbf{x}^{N,t}\}_{t\in T,\, \lambda_N\geq 1}\cup \{\lambda_j\,\overline{\mathbf{p}}^t\cdot\mathbf{x}^{j,t}\}_{ t\in T,\, N>j>i}$ is uniformly bounded away from zero. Therefore, there is $\lambda_i\in (0,1)$ such that for all $j>i$, 
\[f^{j, t}(\mathbf{x}^{j, t})=\phi(\overline{\mathbf{p}}^t\cdot \mathbf{x}^{j, t}, \lambda_j\,\overline{\mathbf{p}}^t\cdot \mathbf{x}^{j, t}) < f^{i, s}(\mathbf{x}^{j, t})=\phi(\overline{\mathbf{p}}^s\cdot \mathbf{x}^{j, t}, \lambda_i\,\overline{\mathbf{p}}^s\cdot \mathbf{x}^{j, t})\]
when $\lambda_i$ is sufficiently small.  Note that $\lambda_i$ can be chosen independently of $\lambda_N$. 

We have shown that there is $\lambda_{N-1}, \ldots, \lambda_2, \lambda_1$ such for any $\lambda_N\geq 1$ and $i, j\in N$ with $i<j$ and $s, t\in T$, $f^{i, s}\not\succsim_{P} f^{j, t}$. We know that $W(\lambda_1,\lambda_2,\ldots,1)\leq 1$, so for $\lambda_N$ sufficiently large,  $W(\lambda_1,\lambda_2,\ldots,\lambda_N)= 1$, as required.  \qed\bigskip

\noindent {\bf Random Utility and Random Price Models.}\,  So far, we have considered discrete cross-sectional data sets. In this section, we show an equivalence between price and preference heterogeneity in a cross-sectional data environment which has continuous probability distributions on expenditure bundles. Let $\mathscr{E}^t=\{\mathbf{e}\in\mathbb{R}^{K}_+: \sum_{k\in K} e_k=m^t\}$ where $m^t$ is the income level. Then, our data set takes the form
\[\mathcal{D}=\{(\mu^t_e, \overline{\mathbf{p}}^t)\}_{t\in T},\]
where $\mu^t_e$ is a probability distribution over $\mathscr{E}^t.$ Let $\mathscr{U}$ be the set of all strictly increasing, strictly concave, and continuous utility functions on $\mathbb{R}^K_{+}$.
We then say that the data set $\mathcal{D}$ is \textbf{RUM-rationalizable} if there is a probability distribution $\rho$ on $\mathscr{U}$ such that for any $t\in T$ and any measurable subset $E\subseteq \mathbb{R}^K_{+}$,
\[\mu^t_e(E)=\rho\big(\{u\in\mathscr{U}: \arg\max_{\mathbf{e}\in \mathscr{E}^t} u(\mathbf{x}(\mathbf{e}, \overline{\mathbf{p}}^t))\in E\}\big).\]
We say that the data set $\mathcal{D}$ is \textbf{RPM-rationalizable} if there is a probability distribution $\eta$ on $\mathbb{R}_{++}$ satisfying $\int \lambda\, d\,\eta(\lambda)=1$ and a utility function $u\in\mathscr{U}$ such that, for any $t\in T$ and any measurable subset $E\subseteq \mathbb{R}^K_{+}$,
\[\mu^t_e(E)=\eta\big(\{\lambda\in\mathbb{R}_{++}: \arg\max_{\mathbf{e}\in \mathscr{E}^t} u\big(\mathbf{x}(\mathbf{e}, \mathbf{p}^{\lambda, t})\big)\in E\}\big),\]
where $\mathbf{p}^{\lambda, t}\equiv \lambda\,\overline{\mathbf{p}}^t$. (Note that our definition of RPM-rationalizability here incorporates a stable price distribution and consistency in expectation w.r.t. the arithmetic mean.)

The next result states that $\cal D$ is RUM-rationalizable if it is RPM-rationalizable; furthermore, whenever $\cal D$ is RUM-rationalizable, then there is another data set $\widehat{\mathcal{D}}$, arbitrarily close to $\cal D$ by the Levy-Prokhorov metric,\footnote{\linespread{1.1} \selectfont For any probability distributions $\mu$ and $\mu'$, the Levy-Prokhorov metric $d_{LP}$ is given by $d_{LP}(\mu, \mu')=\inf\{\epsilon>0: \mu'(E^\epsilon)+\epsilon>\mu(E)\text{ and }\mu(E^\epsilon)+\epsilon>\mu'(E)\text{ for any }E\}$ where $E^\epsilon=\bigcup_{x\in E} \{y\in\mathbb{R}^K:||x-y||<\epsilon\}$.} such that $\widehat{\mathcal{D}}$ is RPM-rationalizable.  In this sense, RUM- and RPM-rationalizability are approximately equivalent.

\begin{theorem}\label{thm-nine} 
Suppose $\mathcal{D}=\{(\mu^t_e, \overline{\mathbf{p}}^t)\}_{t\in T}$ where $\mu^t_e$ are probability distributions which are absolutely continuous with respect to the Lebesgue measure on $\mathscr{E}^t$.\footnote{\linespread{1.1} \selectfont The absolute continuity assumption on $\mu^t_e$ is not strictly necessary but it simplifies the exposition because it guarantees that the intersection of patches has probability measure zero (see \cite{kitamura2018nonparametric}).} If $\mathcal{D}$ is RPM-rationalizable, then it is RUM-rationalizable. Conversely, if $\mathcal{D}$ is RUM-rationalizable, then for any $\epsilon>0$, there are discrete probability distributions $\hat{\mu}^t_e\in\Delta(\mathscr{E}^t)$ such that $\max_{t\in T} d_{LP}(\mu^t_e, \hat{\mu}^t_e)<\epsilon$ and $\widehat{\mathcal{D}}=\{(\hat{\mu}^t_e, \overline{\mathbf{p}}^t)\}_{t\in T}$ is RPM-rationalizable.
\end{theorem}

\noindent \textbf{Proof of Theorem \ref{thm-nine}.}\,  We first show that RPM implies RUM. Suppose $\mathcal{D}$ is RPM-rationalizable; i.e., there is a distribution  $\eta$ on $\mathbb{R}_{++}$ and $u\in\mathscr{U}$ such that, for any $t\in T$ and any measurable subset $E\subseteq \mathbb{R}^K_{+}$, $\mu^t_e(E)=\eta\big(\{\lambda\in\mathbb{R}_{++}: \arg\max_{\mathbf{e}\in \mathscr{E}^t} u\big(\mathbf{x}(\mathbf{e}, \mathbf{p}^{\lambda, t})\big)\in E\}\big)$. For each $\lambda>0$ and $t\in T$, let $\mathbf{e}^{\lambda, t}\equiv\arg\max_{\mathbf{e}\in \mathscr{E}^t} u(\mathbf{x}(\mathbf{e}, \mathbf{p}^{\lambda, t}))$.
Then the data set $\mathcal{O}_{\lambda}=\{(\mathbf{x}(\mathbf{e}^{\lambda, t}, \mathbf{p}^{\lambda, t}), \mathbf{p}^{\lambda, t})\}_{t\in T}$ is rationalized by utility function $u$. Then the data set $\overline{\mathcal{O}}_{\lambda}=\{(\mathbf{x}(\mathbf{e}^{\lambda, t}, \overline{\mathbf{p}}^{t}), \overline{\mathbf{p}}^{t})\}_{t\in T}$ is also rationalizable since $\mathcal{O}_{\lambda}$ obeys GARP if and only if $\overline{\mathcal{O}}_{\lambda}$ obeys GARP (see the proof of Proposition \ref{prop-indistinct}). Hence, there is a utility function $u_{\lambda}$ such that $u_\lambda(\mathbf{x}(\mathbf{e}^{\lambda, t}, \overline{\mathbf{p}}^{t}))\ge u_\lambda(\mathbf{x})\text{ for every }\mathbf{x}\in \mathscr{B}(\overline{\mathbf{p}}^t, m^t)$. Let $\rho$ be a probability distribution over $\mathscr{U}$ such that $\rho(\mathcal{U})=\eta(\{\lambda\in \mathbb{R}_{++}: u_\lambda\in \mathcal{U}\})$ for any measurable subset $\mathcal{U}\in\mathscr{U}$. Therefore, we have $\mu^t_e(E)=\rho\big(\{u_\lambda\in\mathscr{U}: \arg\max_{\mathbf{e}\in \mathscr{E}^t} u_\lambda\big(\mathbf{x}(\mathbf{e}, \overline{\mathbf{p}}^{t})\big)\in E\}\big)$ for any $t\in T$ and any measurable subset $E\subseteq \mathbb{R}^K_{+}$. In other words, $\mathcal{D}$ is RUM-rationalizable.

It remains to show that RUM-rationalizability implies approximate RPM-rationalizability. Our proof uses Theorem 1 of \cite{kitamura2018nonparametric}.  Suppose $\mathcal{D}$ is RUM-rationalizable; i.e., there is $\rho\in\Delta(\mathscr{U})$ such that for any $t\in T$ and any measurable subset $E\subseteq \mathbb{R}^K_{+}$, $\mu^t_e(E)=\rho\big(\{u\in\mathscr{U}: \arg\max_{\mathbf{e}\in \mathscr{E}^t} u(\mathbf{x}(\mathbf{e}, \overline{\mathbf{p}}^t))\in E\}\big)$.

Let $B^t$ be the budget plane at observation $t$, i.e., $B^t=\{\mbx\in \mathbb{R}^{K}_+: \mbx\cdot \overline{\mathbf{p}}^t= m^t\}$. We can find $Y_{1,t}$, $Y_{2,t},\ldots,Y_{L_t,t}$ such that $B^t=\bigcup_{l=1}^{L_t} Y_{l,t}$ where $Y_{l,t}$ has the following property: for any $t'\neq t$, either  $\overline{\mathbf{p}}^{t'} \mathbf{x}-m^{t'}\geq 0$ for all $x\in Y_{l,t}$ or $\overline{\mathbf{p}}^{t'} \mathbf{x}-m^{t'}\leq 0$ for all $x\in Y_{l,t}$.  Following Kitamura and Stoye, we refer to $Y_{l,t}$ as a {\em patch}.   It is clear that two patches on $B^t$ intersect only at their boundaries.

Let $\mu^t$ be the probability measure on $B^t$ induced by the measure $\mu^t_e$ on $\mathscr{E}^t$ and let $\pi_{l, t}=\mu^t(Y_{l, t})$ for each $(l, t)$, i.e., $\pi_{l, t}$ is the probability that a consumption bundle lies in $Y_{l, t}$.  Note that $\sum_{l=1}^{L_t} \pi_{l,t}=1$ since (by the absolute continuity assumption) $\mu^t(Y_{l,t}\bigcap Y_{l',t})=0$ for any $l\neq l'$.   Given any $\epsilon>0$, for each $t\in T$, there is a discrete probability distribution with finite support, $\hat{\mu}_e^t$, on $\mathscr{E}^t$ such that $d_{LP}(\hat{\mu}^t_e, \mu^t_e)<\epsilon$, with $\hat\mu_e^t$ inducing a discrete probability distribution $\hat\mu^t$ on $B^t$  such that $\hat{\mu}^t(Y_{l, t})=\pi_{l, t}$ and $\hat\mu^t(Y_{l,t}\bigcap Y_{l',t})=0$ for any $l\neq l'$.

By Theorem 1 in \cite{kitamura2018nonparametric}, we know that the RUM-rationalizability of $\mathcal{D}$ only depends on the values $\mu^t (Y_{l,t})$ at all patches $Y_{l,t}$.  Since $\hat\mu^t$ satisfies $\hat{\mu}^t(Y_{l, t})=\mu^t(Y_{l, t})$ and $\hat\mu^t(Y_{l,t}\bigcap Y_{l',t})= \mu^t(Y_{l,t}\bigcap Y_{l',t}) =  0$ whenever $l\neq l'$, we conclude that  $\widehat{\cal D}=\{(\hat{\mu}^t_e, \overline{\mathbf{p}}^t)\}_{t\in T}$ is also RUM-rationalizable.  Since $\hat{\mu}^1, \ldots, \hat{\mu}^T$ are discrete probability distributions, there is a discrete probability distribution $\hat{\rho}$ on $\mathscr{U}$ such that, for any $t\in T$ and any measurable subset $E\subseteq \mathbb{R}^K_{+}$, $\hat{\mu}^t_e(E)=\hat{\rho}\big(\{u\in\mathscr{U}: \arg\max_{\mathbf{e}\in \mathscr{E}^t} u(\mathbf{x}(\mathbf{e}, \overline{\mathbf{p}}^t))\in E\}\big)$. Let $\text{supp}(\hat{\rho})=\hat{\mathscr{U}}=\{u^1, \ldots, u^N\}$.

For each $u^i\in \hat{\mathscr{U}}$ and $t\in T$, let $\mathbf{e}^{i, t}=\arg\max_{\mathbf{e}\in \mathscr{E}^t} u^i(\mathbf{x}(\mathbf{e}, \overline{\mathbf{p}}^{t}))$. Then 
$$\mathcal{O}^{i}=\{(\mathbf{x}(\mathbf{e}^{i, t}, \overline{\mathbf{p}}^{t}), \overline{\mathbf{p}}^{t})\}_{t\in T}$$ 
is rationalized by $u^i$. In other words, the data set $\mathcal{E}=\{(\mathbf{e}^{i, t})_{i\in N}, \overline{\mathbf{p}}^{t}\}_{t\in T}$ is rationalizable with heterogeneous preferences. By Theorem \ref{thm-four}, there are real numbers $\lambda_1, \ldots, \lambda_N>0$ such that the stable price distribution ${\cal P}=\{\mathbf{p}^{i,t}\}_{(i,t)\in N\times T}$ with $\mathbf{p}^{i, t}=\lambda_i\, \overline{\mathbf{p}}^t$ rationalizes $\mathcal{E}$ and satisfies the consistency condition $\sum_{i\in N}\lambda_i\,\hat\rho^i=1$ (where $\hat\rho^i$ denotes the probability of $u^i$ under $\hat{\rho}$).  It follows that $\widehat{\cal D}=\{(\hat{\mu}^t_e, \overline{\mathbf{p}}^t)\}_{t\in T}$ is RPM-rationalizable, where the probability of $\lambda_i$ under $\eta$ is $\hat\rho^i$. \hfill \qed

{\footnotesize
\bibliographystyle{ecta}
\bibliography{econref2}

@article{lewbel2017unobserved,
  title={Unobserved preference heterogeneity in demand using generalized random coefficients},
  author={Lewbel, Arthur and Pendakur, Krishna},
  journal={Journal of Political Economy},
  volume={125},
  number={4},
  pages={1100--1148},
  year={2017},
  publisher={University of Chicago Press Chicago, IL}
}

@article{deb2023revealed,
  title={Revealed price preference: theory and empirical analysis},
  author={Deb, Rahul and Kitamura, Yuichi and Quah, John KH and Stoye, J{\"o}rg},
  journal={The Review of Economic Studies},
  volume={90},
  number={2},
  pages={707--743},
  year={2023},
  publisher={Oxford University Press US}
}

@article{mazumdar2005reference,
  title={Reference price research: Review and propositions},
  author={Mazumdar, Tridib and Raj, Sevilimedu P and Sinha, Indrajit},
  journal={Journal of Marketing},
  volume={69},
  number={4},
  pages={84--102},
  year={2005},
  publisher={SAGE Publications Sage CA: Los Angeles, CA}
}

@article{kaplan2019relative,
  title={Relative price dispersion: Evidence and theory},
  author={Kaplan, Greg and Menzio, Guido and Rudanko, Leena and Trachter, Nicholas},
  journal={American Economic Journal: Microeconomics},
  volume={11},
  number={3},
  pages={68--124},
  year={2019},
  publisher={American Economic Association 2014 Broadway, Suite 305, Nashville, TN 37203-2425}
}

@article{van2010has,
  title={Why has house price dispersion gone up?},
  author={Van Nieuwerburgh, Stijn and Weill, Pierre-Olivier},
  journal={The Review of Economic Studies},
  volume={77},
  number={4},
  pages={1567--1606},
  year={2010},
  publisher={Wiley-Blackwell}
}

@article{baye2004price,
  title={Price dispersion in the small and in the large: Evidence from an internet price comparison site},
  author={Baye, Michael R and Morgan, John and Scholten, Patrick},
  journal={The Journal of Industrial Economics},
  volume={52},
  number={4},
  pages={463--496},
  year={2004},
  publisher={Wiley Online Library}
}

@article{brown2002does,
  title={Does the Internet make markets more competitive? Evidence from the life insurance industry},
  author={Brown, Jeffrey R and Goolsbee, Austan},
  journal={Journal of Political Economy},
  volume={110},
  number={3},
  pages={481--507},
  year={2002},
  publisher={The University of Chicago Press}
}

@article{sorensen2000equilibrium,
  title={Equilibrium price dispersion in retail markets for prescription drugs},
  author={Sorensen, Alan T},
  journal={Journal of Political Economy},
  volume={108},
  number={4},
  pages={833--850},
  year={2000},
  publisher={The University of Chicago Press}
}

@article{stigler1961economics,
  title={The economics of information},
  author={Stigler, George J},
  journal={Journal of Political Economy},
  volume={69},
  number={3},
  pages={213--225},
  year={1961},
  publisher={The University of Chicago Press}
}

@book{prais1971analysis,
  title={The analysis of family budgets},
  author={Prais, Sigbert Jon and Houthakker, Hendrik S},
  volume={4},
  year={1971},
  publisher={CUP Archive}
}

@article{dierker1984price,
  title={Price-dispersed preferences and C1 mean demand},
  author={Dierker, Egbert and Dierker, Hildegard and Trockel, Walter},
  journal={Journal of Mathematical Economics},
  volume={13},
  number={1},
  pages={11--42},
  year={1984},
  publisher={Elsevier}
}

@article{mas1977some,
  title={Some generic properties of aggregate excess demand and an application},
  author={Mas-Colell, Andreu and Neuefeind, Wilhelm},
  journal={Econometrica},
  pages={591--599},
  year={1977},
  publisher={JSTOR}
}

@article{heidhues2008competition,
  title={Competition and price variation when consumers are loss averse},
  author={Heidhues, Paul and K{\H{o}}szegi, Botond},
  journal={American Economic Review},
  volume={98},
  number={4},
  pages={1245--68},
  year={2008}
}

@article{brown2007nonparametric,
  title={The nonparametric approach to applied welfare analysis},
  author={Brown, Donald J and Calsamiglia, Caterina},
  journal={Economic Theory},
  volume={31},
  number={1},
  pages={183--188},
  year={2007},
  publisher={Springer}
}

@article{muellbauer1980estimation,
  title={The estimation of the Prais-Houthakker model of equivalence scales},
  author={Muellbauer, John},
  journal={Econometrica},
  pages={153--176},
  year={1980},
  publisher={JSTOR}
}

@article{grandmont1987distributions,
  title={Distributions of preferences and the ``Law of Demand"},
  author={Grandmont, Jean-Michel},
  journal={Econometrica},
  pages={155--161},
  year={1987},
  publisher={JSTOR}
}

@article{barten1964consumer,
  title={Consumer demand functions under conditions of almost additive preferences},
  author={Barten, Anton P},
  journal={Econometrica},
  pages={1--38},
  year={1964},
  publisher={JSTOR}
}

@article{varian1988revealed,
  title={Revealed preference with a subset of goods},
  author={Varian, Hal R},
  journal={Journal of Economic Theory},
  volume={46},
  number={1},
  pages={179--185},
  year={1988},
  publisher={Elsevier}
}

@article{cattaneo2020random,
  title={A random attention model},
  author={Cattaneo, Matias D and Ma, Xinwei and Masatlioglu, Yusufcan and Suleymanov, Elchin},
  journal={Journal of Political Economy},
  volume={128},
  number={7},
  pages={2796--2836},
  year={2020},
  publisher={The University of Chicago Press Chicago, IL}
}

@article{dellavigna2006paying,
  title={Paying not to go to the gym},
  author={DellaVigna, Stefano and Malmendier, Ulrike},
  journal={American Economic Review},
  volume={96},
  number={3},
  pages={694--719},
  year={2006}
}

@article{gabaix2006shrouded,
  title={Shrouded attributes, consumer myopia, and information suppression in competitive markets},
  author={Gabaix, Xavier and Laibson, David},
  journal={The Quarterly Journal of Economics},
  volume={121},
  number={2},
  pages={505--540},
  year={2006},
  publisher={MIT Press}
}

@article{kitamura2018nonparametric,
  title={Nonparametric analysis of random utility models},
  author={Kitamura, Yuichi and Stoye, J{\"o}rg},
  journal={Econometrica},
  volume={86},
  number={6},
  pages={1883--1909},
  year={2018},
  publisher={Wiley Online Library}
}

@article{matvejka2015rigid,
  title={Rigid pricing and rationally inattentive consumer},
  author={Mat{\v{e}}jka, Filip},
  journal={Journal of Economic Theory},
  volume={158},
  pages={656--678},
  year={2015},
  publisher={Elsevier}
}

@article{brown2010shrouded,
  title={Shrouded attributes and information suppression: Evidence from the field},
  author={Brown, Jennifer and Hossain, Tanjim and Morgan, John},
  journal={The Quarterly Journal of Economics},
  volume={125},
  number={2},
  pages={859--876},
  year={2010},
  publisher={MIT Press}
}

@article{gabaix2014sparsity,
  title={A sparsity-based model of bounded rationality},
  author={Gabaix, Xavier},
  journal={The Quarterly Journal of Economics},
  volume={129},
  number={4},
  pages={1661--1710},
  year={2014},
  publisher={MIT Press}
}

@article{chetty2009salience,
  title={Salience and taxation: Theory and evidence},
  author={Chetty, Raj and Looney, Adam and Kroft, Kory},
  journal={American Economic Review},
  volume={99},
  number={4},
  pages={1145--77},
  year={2009}
}

@article{brown1996testable,
  title={Testable restrictions on the equilibrium manifold},
  author={Brown, Donald J and Matzkin, Rosa L},
  journal={Econometrica},
  pages={1249--1262},
  year={1996},
  publisher={JSTOR}
}

@article{mcfadden1973conditional,
  title={Conditional logit analysis of qualitative choice behavior},
  author={McFadden, D},
  journal={Frontiers in Econometrics},
  pages={105--142},
  year={1973},
  publisher={Academic press}
}

@article{Block1960,
  title={Random Orderings and Stochastic Theories of Responses},
  author={Block, D and Marshak, J},
  journal={Contributions to Probability and Statistics, ed. by I Olkin et al. Stanford},
  pages={2381--2391},
  year={1960},
  publisher={Stanford University Press}
}

@article{quah1997law,
  title={The law of demand when income is price dependent},
  author={Quah, John K-H},
  journal={Econometrica},
  pages={1421--1442},
  year={1997},
  publisher={JSTOR}
}

@article{grandmont1992transformations,
  title={Transformations of the commodity space, behavioral heterogeneity, and the aggregation problem},
  author={Grandmont, Jean-Michel},
  journal={Journal of Economic Theory},
  volume={57},
  number={1},
  pages={1--35},
  year={1992},
  publisher={Elsevier}
}

@article{blundell1991information,
  title={The information content of equivalence scales},
  author={Blundell, Richard and Lewbel, Arthur},
  journal={Journal of Econometrics},
  volume={50},
  number={1-2},
  pages={49--68},
  year={1991},
  publisher={North-Holland}
}

@book{deaton1997analysis,
  title={The analysis of household surveys: a microeconometric approach to development policy},
  author={Deaton, Angus},
  year={1997},
  publisher={The World Bank}
}

@article{lewbel1989household,
  title={Household equivalence scales and welfare comparisons},
  author={Lewbel, Arthur},
  journal={Journal of Public Economics},
  volume={39},
  number={3},
  pages={377--391},
  year={1989},
  publisher={Elsevier}
}

@article{varian1982nonparametric,
  title={The nonparametric approach to demand analysis},
  author={Varian, Hal R},
  journal={Econometrica},
  pages={945--973},
  year={1982},
  publisher={JSTOR}
}

@article{afriat1967construction,
  title={The construction of utility functions from expenditure data},
  author={Afriat, Sydney N},
  journal={International Economic Review},
  volume={8},
  number={1},
  pages={67--77},
  year={1967},
  publisher={JSTOR}
}

@article{cerreia2019deliberately,
  title={Deliberately stochastic},
  author={Cerreia-Vioglio, Simone and Dillenberger, David and Ortoleva, Pietro and Riella, Gil},
  journal={American Economic Review},
  volume={109},
  number={7},
  pages={2425--45},
  year={2019}
}

@article{masatlioglu2012revealed,
	Author = {Masatlioglu, Yusufcan and Nakajima, Daisuke and Ozbay, Erkut Y},
	Journal = {The American Economic Review},
	Number = {5},
	Pages = {2183--2205},
	Publisher = {American Economic Association},
	Title = {Revealed attention},
	Volume = {102},
	Year = {2012}}

@article{manzini2014stochastic,
	Author = {Manzini, Paola and Mariotti, Marco},
	Journal = {Econometrica},
	Number = {3},
	Pages = {1153--1176},
	Publisher = {Wiley Online Library},
	Title = {Stochastic choice and consideration sets},
	Volume = {82},
	Year = {2014}}

@book{luce1959individual,
	Author = {Luce, R Duncan},
	Publisher = {John Wiley and sons},
	Title = {Individual choice behavior: A theoretical analysis},
	Year = {1959}}

@article{gennaioli2013salience,
	Author = {Bordalo, Pedro and Gennaioli, Nicola and Shleifer, Andrei},
	Journal = {Journal of Political Economy},
	Number = {5},
	Pages = {803--843},
	Title = {Salience and consumer choice},
	Volume = {121},
	Year = {2013}}

@article{koszegi2006model,
	Author = {K{\H{o}}szegi, Botond and Rabin, Matthew},
	Journal = {The Quarterly Journal of Economics},
	Number = {4},
	Title = {A model of reference-dependent preferences},
	Volume = {121},
	Year = {2006}}

@article{koszegi2007reference,
	Author = {K{\H{o}}szegi, Botond and Rabin, Matthew},
	Journal = {American Economic Review},
	Number = {4},
	Pages = {1047--1073},
	Publisher = {American Economic Association},
	Title = {Reference-dependent risk attitudes},
	Volume = {97},
	Year = {2007}}

@article{fudenberg2015stochastic,
	Author = {Fudenberg, Drew and Iijima, Ryota and Strzalecki, Tomasz},
	Journal = {Econometrica},
	Number = {6},
	Pages = {2371--2409},
	Publisher = {Wiley Online Library},
	Title = {Stochastic choice and revealed perturbed utility},
	Volume = {83},
	Year = {2015}}

@article{falmagne1978representation,
	Author = {Falmagne, Jean-Claude},
	Journal = {Journal of Mathematical Psychology},
	Number = {1},
	Pages = {52--72},
	Publisher = {Elsevier},
	Title = {A representation theorem for finite random scale systems},
	Volume = {18},
	Year = {1978}}

@article{mcfadden1990stochastic,
	Author = {McFadden, Daniel and Richter, Marcel K},
	Journal = {Preferences, Uncertainty, and Optimality, Essays in Honor of Leo Hurwicz, Westview Press: Boulder, CO},
	Pages = {161--186},
	Title = {Stochastic rationality and revealed stochastic preference},
	Year = {1990}}

@article{brady2016menu,
	Author = {Brady, Richard L and Rehbeck, John},
	Journal = {Econometrica},
	Number = {3},
	Pages = {1203--1223},
	Publisher = {Wiley Online Library},
	Title = {Menu-Dependent Stochastic Feasibility},
	Volume = {84},
	Year = {2016}}
}

\end{document}